\documentclass[twocolumn]{aastex701}
\usepackage{apjfonts} 

\newif\preprint 
\usepackage{amsmath,amstext}
\usepackage[T1]{fontenc}
\usepackage{comment}
\usepackage{natbib}
\usepackage{multirow}
\usepackage{tabularx}
\usepackage{array}
\usepackage{verbatim}
\usepackage{hyperref}
\usepackage{tikz}
\hypersetup{
    colorlinks=true,
    linkcolor=blue,
    filecolor=magenta,      
    urlcolor=cyan,
  }
  \usepackage{footmisc}
\graphicspath{{./}{figures/}}

\begin{document}
\def\lesssim{\mathrel{\hbox{\rlap{\hbox{%
 \lower4pt\hbox{$\sim$}}}\hbox{$<$}}}}
\def\gtrsim{\mathrel{\hbox{\rlap{\hbox{%
 \lower4pt\hbox{$\sim$}}}\hbox{$>$}}}}
\def\I.{\kern.2em{\sc i}} 
\def\II.{\kern.2em{\sc ii}} 
\def\III.{\kern.2em{\sc iii}} 
\def\IV.{\kern.2em{\sc iv}} 
\def\arcs{\hbox{$^{\prime\prime}$}}

\newcommand{\hst}{{HST}}
\newcommand{\spitzer}{{Spitzer}}
\newcommand{\chandra}{{Chandra}}
\newcommand{\sirtf}{{Spitzer}}
\newcommand{\cxo}{{CXO}}
\newcommand{\hdf}{HDF--N}
\newcommand{\wfu}{\hbox{$U_{300}$}}
\newcommand{\wfb}{\hbox{$B_{450}$}}
\newcommand{\wfv}{\hbox{$V_{606}$}}
\newcommand{\wfi}{\hbox{$I_{814}$}}
\newcommand{\acsv}{\hbox{$V_{606}$}}
\newcommand{\acsi}{\hbox{$I_{814}$}}
\newcommand{\nicj}{\hbox{$J_{110}$}}
\newcommand{\nich}{\hbox{$H_{160}$}}
\newcommand{\mirifive}{\hbox{$[5.6]$}}
\newcommand{\miriseven}{\hbox{$[7.7]$}}
\newcommand{\mirirten}{\hbox{$[10]$}}
\newcommand{\ab}{{\rm AB}}
\newcommand{\vega}{{\rm Vega}}
\newcommand{\ks}{\hbox{$K_s$}}
\newcommand{\tfit}{\hbox{TFIT}}
\newcommand{\sigmarms}{\sigma_\mathrm{rms}}
\newcommand{\gsim}{\gtrsim}
\newcommand{\lsim}{\lesssim}
\newcommand{\etal}{et al.}
\newcommand{\eg}{e.g.}
\newcommand{\ie}{i.e.}
\newcommand{\hbeta}{\hbox{H$\beta$}}
\newcommand{\he}{\hbox{H$\epsilon$}}
\newcommand{\hd}{\hbox{H$\delta$}}
\newcommand{\hg}{\hbox{H$\gamma$}}
\newcommand{\hgamma}{\hbox{H$\gamma$}}
\newcommand{\hc}{\hbox{H$\gamma$}}
\newcommand{\hb}{\hbox{H$\beta$}}
\newcommand{\ha}{\hbox{H$\alpha$}}
\newcommand{\oiii}{\hbox{[\ion{O}{3}]}}
\newcommand{\oiiiuv}{\hbox{\ion{O}{3}]}}
\newcommand{\civ}{\hbox{\ion{C}{4}}}
\newcommand{\ciii}{\hbox{\ion{C}{3}]}}
\newcommand{\heii}{\hbox{\ion{He}{2}}}
\newcommand{\hei}{\hbox{\ion{He}{1}}}
\newcommand{\aliii}{\hbox{\ion{Al}{3}}}
\newcommand{\siiv}{\hbox{\ion{Si}{4}}}
\newcommand{\oiv}{\hbox{\ion{O}{4}]}}
\newcommand{\neiii}{\hbox{[\ion{Ne}{3}]}}
\newcommand{\lstar}{\hbox{$L^\ast$}}
\newcommand{\extinctA}{\hbox{$A_{1700}$}}
\newcommand{\mathS}{\hbox{$\mathcal{S}$}}
\newcommand{\mathR}{\hbox{$\mathcal{R}$}}
\newcommand{\NLBG}{33} 
\newcommand{\dchi}{\hbox{$\Delta \chi^2$}}
\newcommand{\ebv}{\hbox{$E(B-V)$}}
\newcommand{\royalsoceity}{Phil.\ Trans.\ R.\ Soc.\ Lond.\ A}

\newcommand{\mstar}{\hbox{$M_\ast$}}
\newcommand{\Mstar}{\hbox{$M_\ast$}}

\newcommand{\zstar}{\hbox{$Z_\ast$}}
\newcommand{\Zstar}{\hbox{$Z_\ast$}}

\newcommand{\Msol}{\hbox{$M_\odot$}}
\newcommand{\Zsol}{\hbox{$Z_\odot$}}
\newcommand{\msol}{\hbox{$M_\odot$}}
\newcommand{\zsol}{\hbox{$Z_\odot$}}
\newcommand{\fion}[2]{\hbox{[\ion#1#2]}}
\newcommand{\infinity}{\hbox{$\infty$}}
\newcommand{\efolding}{$e$--folding}
\newcommand{\lya}{Lyman~$\alpha$}
\newcommand{\qso}{{Q0122+0338}}
\newcommand{\fos}{{FOS}}
\newcommand{\za}{\hbox{$z_\mathrm{a}$}}
\newcommand{\ze}{\hbox{$z_\mathrm{e}$}}
\newcommand{\NH}{\hbox{$N(\mathrm{H})$}}
\newcommand{\phc}{\phm{:}}
\newcommand{\inprep}{\textit{in prep}}
\newcommand{\distf}{\hbox{$f(M,\dot M,t)$}}
\newcommand{\degree}{\hbox{$^\circ$}}
\newcommand{\mydot}{\hbox{$\bullet$}}
\newcommand\myRoman[1]{\@Roman{#1}\relax}%
\newcommand{\fourge}{{\sc FourGE}}
\newcommand{\sfrten}{\hbox{SFR$_\mathrm{10}$}}
\newcommand{\sfrcen}{\hbox{SFR$_\mathrm{100}$}}

\newcommand{\mone}{\hbox{$[3.6]$}}
\newcommand{\mtwo}{\hbox{$[4.5]$}}
\renewcommand{\plotfiddle}[7]{
        \centering 
        \leavevmode
        \vbox to#2{\rule{0pt}{#2}}
        \includegraphics{#1}
}
\newcommand{\zfourge}{\hbox{ZFOURGE}}

\newcommand{\zftwo}{ZFK2}

\newcommand{\herschel}{{Herschel}}
\newcommand{\jwst}{{JWST}}
\newcommand{\kb}{\hbox{$K_b$}}
\newcommand{\kr}{\hbox{$K_r$}}
\newcommand{\jone}{\hbox{$J_1$}}
\newcommand{\jtwo}{\hbox{$J_2$}}
\newcommand{\jthree}{\hbox{$J_3$}}
\newcommand{\hs}{\hbox{$H_s$}}
\newcommand{\hl}{\hbox{$H_l$}}
\newcommand{\lir}{\hbox{$L_\mathrm{IR}$}}

\newcommand{\wfcj}{\hbox{$J_{125}$}}
\newcommand{\wfch}{\hbox{$H_{160}$}}
\newcommand{\OH}{\hbox{$12 + \log(\mathrm{O/H})$}}

\newcommand{\myhref}[1]{\href{#1}{#1}}

\newcommand{\lcdm}{$\Lambda$CDM}
\newcommand{\LCDM}{$\Lambda$CDM}
\renewcommand{\mone}{\hbox{$[3.6]$}}
\renewcommand{\mtwo}{\hbox{$[4.5]$}}
\newcommand{\mthree}{\hbox{$[5.8]$}}
\newcommand{\mfour}{\hbox{$[8.0]$}}
\newcommand{\reff}{\hbox{$r_\mathrm{eff}$}}
\newcommand{\sersic}{S\'ersic}
\newcommand{\um}{\hbox{$\mu$m}}
\newcommand{\pstart}[1]{\noindent \textbf{#1}}
\definecolor{aggiemaroon}{HTML}{500000}

\renewcommand{\neiii}{\hbox{[Ne\,{\sc iii}]}}
\newcommand{\oii}{\hbox{[O\,{\sc ii}]}}
\renewcommand{\oiii}{\hbox{[O\,{\sc iii}]}}
\newcommand{\nii}{\hbox{[N\,{\sc ii}]}}
\newcommand{\sii}{\hbox{[S\,{\sc ii}]}}
\renewcommand{\siiv}{\hbox{Si\,{\sc iv}}}
\newcommand{\hii}{\hbox{H\,{\sc ii}}}
\renewcommand{\heii}{\hbox{He\,{\sc ii}}}
\newcommand{\llambda}{\lambda\lambda}
\newcommand{\myshrink}{\vspace{-7pt}}
\newcommand{\mylinebreak}{\vspace{8pt}}

\newcommand{\grizli}{\hbox{\texttt{grizli}}}
\newcommand{\linmixi}{\hbox{\texttt{LINMIX}}}
\newcommand{\izi}{\hbox{\texttt{IZI}}}
\newcommand{\ppxf}{\hbox{\texttt{pPXF}}}
\newcommand{\bagpipes}{\hbox{\texttt{BAGPIPES}}}
\newcommand{\sextractor}{\hbox{\texttt{SE}}}

\newcommand{\fesc}{\hbox{$f_\mathrm{esc}$}}
\newcommand{\xiion}{\hbox{$\xi_\mathrm{ion}$}}
\newcommand{\luv}{\hbox{$L_\mathrm{UV}$}}
\newcommand{\muv}{\hbox{$M_\mathrm{UV}$}}
\newcommand{\ndotion}{\hbox{$\dot n_\mathrm{ion}$}}

\definecolor{aggiemaroon}{HTML}{500000}

\def\vshiftfig#1#2#3#4{\hfill 
\vbox{\parskip=0pt\hsize=#2
\vskip{#4} \includegraphics[width=#2]{#1}\vskip2pt
\vtop{\centering
\footnotesize
\hsize=#2
#3\vskip1pt
}}\hfill}

\newcommand{\todo}[1]{\textcolor{aggiemaroon}{\tt #1}}

\title{\large \bf Implications of Broad \oiii\ $\lambda$4364 and UV Line Emission in Two Little Red Dots at $\mathbf{z\approx7-8}$}

\correspondingauthor{Casey Papovich}
\email{papovich@tamu.edu}

\author[0000-0001-7503-8482]{Casey Papovich}
\affiliation{Department of Physics and Astronomy, Texas A\&M University, College
Station, TX, 77843-4242 USA}
\affiliation{George P.\ and Cynthia Woods Mitchell Institute for
 Fundamental Physics and Astronomy, Texas A\&M University, College
 Station, TX, 77843-4242 USA}
 \email{papovich@tamu.edu}

\author[0000-0001-5749-5452]{Kaila Ronayne}
\affiliation{Department of Physics and Astronomy, Texas A\&M University, College Station, TX, 77843-4242 USA}
\affiliation{George P.\ and Cynthia Woods Mitchell Institute for Fundamental Physics and Astronomy, Texas A\&M University, College Station, TX, 77843-4242 USA}
\email{kaila\_ronayne@tamu.edu}

  \author[0000-0003-3424-3230]{Weida Hu}
\affiliation{Department of Physics and Astronomy, Texas A\&M
  University, College Station, TX, 77843-4242 USA}
\affiliation{George P.\ and Cynthia Woods Mitchell Institute for
  Fundamental Physics and Astronomy, Texas A\&M University, College
  Station, TX, 77843-4242 USA}
 \email{weidahu@tamu.edu}

\author[0000-0002-8360-3880]{Dale D. Kocevski}
\affiliation{Department of Physics and Astronomy, Colby College, Waterville, ME 04901, USA}
\email{dkocevsk@colby.edu}

\author[0000-0000-0000-0000]{Pablo Arrabal Haro}
\altaffiliation{NASA Postdoctoral Fellow}
\affiliation{Astrophysics Science Division, NASA Goddard Space Flight Center, 8800 Greenbelt Rd, Greenbelt, MD 20771, USA}
\email{pablo.arrabalharo@nasa.gov}

 \author[0000-0001-6251-4988]{Taylor A. Hutchison}
\altaffiliation{NASA Postdoctoral Fellow}
\affiliation{Astrophysics Science Division, NASA Goddard Space Flight Center, 8800 Greenbelt Rd, Greenbelt, MD 20771, USA}
\email{taylor.hutchison@nasa.gov}

\author[0000-0001-8519-1130]{Steven L. Finkelstein}
\affiliation{Department of Astronomy, The University of Texas at Austin, Austin, TX 78712, USA}
\affiliation{Cosmic Frontier Center, The University of Texas at Austin, Austin, TX 78712, USA} 
\email{stevenf@astro.as.utexas.edu}

\author[orcid=0000-0003-3216-7190,sname='Lambrides']{Erini Lambrides}
\affiliation{Astrophysics Science Division, NASA Goddard Space Flight Center, 8800 Greenbelt Rd, Greenbelt, MD 20771, USA}
\affiliation{Department of Astronomy, University of Maryland, College Park, MD 20742, USA}
\affiliation{Center for Research and Exploration in Space Science and Technology, NASA/GSFC, Greenbelt, MD 20771 USA}
\email{erini.lambrides@nasa.gov}

\author[0000-0003-2366-8858]{Rebecca L. Larson}
\altaffiliation{Giacconi Postdoctoral Fellow}
\affil{Space Telescope Science Institute, 3700 San Martin Drive, Baltimore, MD 21218, USA}
\email{rlarson@stsci.edu}

\author[0000-0001-8534-7502]{Bren E. Backhaus}
\affil{Department of Physics and Astronomy, University of Kansas, Lawrence, KS 66045, USA}
\email{bren.backhaus@ku.edu}

\author[0000-0002-9921-9218]{Micaela B. Bagley}
\affil{Southeastern Universities Research Association, Washington, DC 20005, USA}
\affiliation{Astrophysics Science Division, NASA Goddard Space Flight Center, 8800 Greenbelt Rd, Greenbelt, MD 20771, USA}
\affiliation{Department of Astronomy, The University of Texas at Austin, Austin, TX 78712, USA}
\email{micaela.bagley@gmail.com}

\author[0000-0001-7151-009X]{Nikko J. Cleri}
\affiliation{Department of Astronomy and Astrophysics, The Pennsylvania State University, University Park, PA 16802, USA}
\affiliation{Institute for Computational and Data Sciences, The Pennsylvania State University, University Park, PA 16802, USA}
\affiliation{Institute for Gravitation and the Cosmos, The Pennsylvania State University, University Park, PA 16802, USA}
\email{cleri@psu.edu}

\author[0000-0001-5414-5131]{Mark Dickinson}
\affiliation{NSF NOIRLab, 950 N. Cherry Ave., Tucson, AZ 85719, USA}
 \email{mark.dickinson@noirlab.edu}

\author[0000-0002-9426-7456,gname=Ray,sname=Garner,suffix=III]{Ray Garner, III}
\affiliation{Department of Physics and Astronomy, Texas A\&M University, College Station, TX, 77843-4242 USA}
\affiliation{George P.\ and Cynthia Woods Mitchell Institute for  Fundamental Physics and Astronomy, Texas A\&M University, College  Station, TX, 77843-4242 USA}
 \email{ray.three.garner@gmail.com}

\author[0000-0001-9187-3605]{Jeyhan S. Kartaltepe}
\affiliation{Laboratory for Multiwavelength Astrophysics, School of Physics and Astronomy, Rochester Institute of Technology, 84 Lomb Memorial Drive, Rochester, NY 14623, USA}
\email{jeyhan@astro.rit.edu}

\author[0000-0002-5588-9156]{Vasily Kokorev}
\affiliation{Department of Astronomy, The University of Texas at Austin, Austin, TX 78712, USA}
\affiliation{Cosmic Frontier Center, The University of Texas at Austin, Austin, TX 78712, USA} 
\email{vasily.kokorev.astro@gmail.com}

\author[0000-0002-4606-4240]{Grace M. Olivier}
\affiliation{The Observatories of the Carnegie Institution for Science, 813 Santa Barbara Street, Pasadena, CA 91101, USA}
\email{golivier@carnegiescience.edu}

\author[0000-0003-1282-7454]{Anthony J. Taylor}
\email{anthony.taylor@austin.utexas.edu}
\affiliation{Department of Astronomy, The University of Texas at Austin, Austin, TX 78712, USA}
\affiliation{Cosmic Frontier Center, The University of Texas at Austin, Austin, TX 78712, USA} 

\author[0000-0002-1410-0470]{Jonathan R. Trump}
\affiliation{Department of Physics, 196A Auditorium Road, Unit 3046, University of Connecticut, Storrs, CT 06269, USA}
\email{jonathan.trump@uconn.edu}

\author[0000-0001-7593-9205]{Jiayang Yang}\affiliation{Department of Physics and Astronomy, Texas A\&M University, College Station, TX, 77843-4242 USA}\affiliation{George P.\ and Cynthia Woods Mitchell Institute for
 Fundamental Physics and Astronomy, Texas A\&M University, College Station, TX, 77843-4242 USA}\email{annabellayang@tamu.edu}


\begin{abstract}
We present deep, NIRSpec G140M and G395M spectroscopy of Little Red Dots (LRDs) at $z=6.68$ and $z=8.35$.  Both LRDs show broad Balmer and \oiii\ $\lambda$4364 emission. The broad \oiii\ $\lambda$4364 lines have FWHM~$\simeq$1000~km s$^{-1}$, about 1/3 that of the \hbeta\ lines. Assuming gas temperatures $T_e\sim$15,000--25,000~K, the \oiii\ $\lambda$4364/\oiii\ $\lambda$5008 ratios of the broad lines yield high gas densities, $\log n/\mathrm{cm^{-3}} = 6.3$ to 7.9, 3--10$\times$ higher than those in broad-line regions of low-redshift quasars.   If the broad-lines trace virial motions, it is evidence for metal-enhanced gas clouds, $\sim$1-10~pc from the LRD engine.  Both LRDs show narrow [\ion{C}{3}] $\lambda$1907 + \ciii\ $\lambda$1909, and \oiiiuv\ $\lambda\lambda$1661,1666. The \ciii\ ratios yield narrow-line gas densities, $\log n_e/\mathrm{cm^{-3}} = 4.2-5.2$, similar to those in other star-forming galaxies.   The line equivalent widths, EW(\oiiiuv), EW(\ciii), are at, or exceed, limits expected for stellar populations,  likely requiring an additional ionizing source.  The LRDs also have \oiii\ $\lambda$4364/\hgamma\ ratios that favor ionization from an accretion disk, possibly combined with stars.  Both LRDs show nitrogen enhancement based on detections of \ion{N}{3}] $\lambda$1746 or \ion{N}{4}] $\lambda$1486, which may imply rapid, recent star-formation.   These results favor a scenario where the LRD gas envelopes are highly stratified, having high-density clouds with a non-unity covering factors and a complex geometry, such that ionizing radiation from the LRD accretion disk, combined with that from star-forming regions, produce the nebular emission features. 
%
%
\end{abstract}

\section{Introduction}\label{section:introduction}

Little Red Dots (LRDs) are a population of luminous, compact, and red objects discovered by \jwst\ and primarily denizens of the distant universe (e.g., \citealt{Kocevski_2023,Matthee_2024}).   LRDs exhibit blue rest-UV and red optical--to--near-IR colors, giving rise to ``V''--shaped spectral energy distributions (SEDs, \citealt{Kocevski_2025}).     The rest-frame optical/near-IR SEDs of LRDs appear similar to thermal sources with characteristic effective temperatures, $T_e \sim 5000$~K \citep{deGraaff_2025b, Lin_2026, Sun_2026,Umeda_2026}.   Their compact morphologies imply sizes of $\lesssim 30-100$~pc \citep{Baggen_2024, Furtak_2024, Guia_2024}.  These properties make LRDs a novel type of source. 

The combined LRD properties favor scenarios where they are powered by intermediate-- or supermassive--black-holes (SMBHs) in a transitory, rapid growth phase surrounded by gas cocoons \citep{Naidu_2025,Begelman_2026,Asada_2026}, though other possibilities involving stellar-origins exist \citep[e.g.,][]{Baggen_2024,Chisholm_2026}.  This picture is complicated as LRDs lack indications commonly associated with obscured active galactic nuclei (AGN), including an absence of significant dust reddening based on mid-IR, far-IR, and radio observations \citep[e.g.,][]{Akins_2024,Casey_2024,Gloudemans_2025,Setton_2025,Ronayne_2026}, and a lack of X-ray emission \citep{Ananna_2024, Yue_2024, Lambrides_2026_b}.  Therefore, if LRDs are a class of AGN, they represent a physical scenario that is not understood.  Furthermore, LRDs appear to be a phenomenon primarily of the high-redshift universe, with a number density that increases with redshift, exceeding that of quasars at $z > 4$ \citep{Akins_2024,kokorev_2024, Matthee_2024,Juodzbalis_2025,Kocevski_2025}.  As LRDs likely represent a crucial evolutionary phase of SMBHs, understanding their nature is paramount. 

Spectroscopy of LRDs imply dense gas surrounds the ionizing source, but the details are difficult to interpret.  The spectra show characteristic red continua with Balmer absorption, indicating warm, dense partially ionized gas \citep{ deGraaff2025a,Inayoshi_2025,Ji_X_2025,Juodzbalis_2025, Naidu_2025}.   The spectra also show broad hydrogen and \ion{He}{1} emission lines \citep{Wang_b_2025,Kokorev_2026} with widths of $\simeq$1000--3000~km s$^{-1}$, superimposed with blue-shifted and red-shifted absorption features indicative of warm--gas flows, and/or rapid Keplerian motions in the vicinity of the central engine.  Some LRDs show evidence of exponential broad-line profiles \citep{Juodzbalis_2025}, which indicates electron-scattering of photons may drive the wings of the broad emission lines rather than Doppler motions \citep{Rusakov_2026} or the presence of a stratified broad-line-emitting gas region \citep{Madau_2026c}.  It may also be that LRDs represent a population with an intrinsic range of properties, complicating the interpretation \cite{Asada_2026,Ronayne_2026}.

Recently, studies using radiative transfer show that the SEDs, Balmer and \ion{He}{1}, features of LRDs can be reproduced with ``dense-gas'' models that have high column densities of dense gas surrounding a central, accreting SMBH \citep{Taylor_2025,Ronayne_2026}.  These models include  a ``black-hole--star'' (BH*) variant \citep{Naidu_2025} and ``quasi-star'' variant \citep{Begelman_2008}, where the SMBH is completely enveloped by thermalized gas.    This has the ability to account for the shape of the rest-optical--to--near-IR continuum, the Balmer and \ion{He}{1} absorption, and lack of X-ray detections \citealt{Yue_2024, Maiolino_2025}, without the need for significant dust attenuation \citep{Mazzolari_2024,Gloudemans_2025,Setton_2025}. 

While encouraging, these dense--gas models leave many unaddressed problems and face other challenges \citep[e.g.,][]{Asada_2026}.   One problem with these models is that spectroscopy of rest-optical emission lines of LRDs are typically more consistent with photoionization by \ion{H}{2} regions compared to expectations from AGN \cite[e.g.,][] {Kocevski_2023,Harikane_2023,Maiolino_2024,Ubler_2024,Juodzbalis_2025,Wang_b_2025}.   LRDs show strong narrow metal lines, including \oii\ $\lambda\lambda$3727,3730, [\ion{Ne}{3}] $\lambda$3870, \oiii\ $\lambda\lambda$4960,5008, none of which are produced by the dense-gas models.  Indeed, some dense-gas models require unity covering fractions, such that no hard ionizing radiation from the central accretion disk escape \citep{Naidu_2025}.  In these models, the ionizing radiation comes from central star clusters \citep[e.g.,][]{Asada_2026}.   Therefore if the metal lines show indications of being illuminated by the ionizing radiation of the LRD engine, this would favor non-unity covering fractions and provide information about any AGN accretion disk.  Some LRDs show excess \ion{He}{2} $\lambda$4687/\hbeta\ ratios, indicative of this scenario \citep{Wang_2026}.  

Another way to test the ionizing source is to look for indications of coronal lines, directly tracing hard ionizing photons, $E \gtrsim 100$~eV, from an accretion disk \citep[e.g.,][]{Cleri_2023a,Cleri_2023b,Cleri_2025}, which has been detected in low-redshift analogs of LRDs \citep[e.g.,][]{Casey_2026}.   Such lines, including [\ion{Fe}{7}] $\lambda$5160, and [\ion{Fe}{10}] $\lambda$6376, have possibly been detected in some LRDs \citep{Kocevski_2025,Lambrides_2025}, but this is not yet robust and may be contaminated by [\ion{Fe}{2}] line emission \citep{Torralba_2026}.  

Lastly,  rest-frame UV lines can provide a test of the ionizing source.  Many of these lines are very sensitive to the gas density and ionization state \citep{Jaskot_2016,Feltre_2016,Hutchison_2019,Hirschmann_2023}. However, to date, studies of the rest-UV lines in LRDs stem only from low-resolution \jwst/NIRSpec prism observations \cite[e.g.,][]{Ji_X_2026}, which have not provided conclusive results. 

\begin{figure*}
\begin{center}
     \includegraphics[trim={0pt 0 0 0}, clip, width=0.9\textwidth]{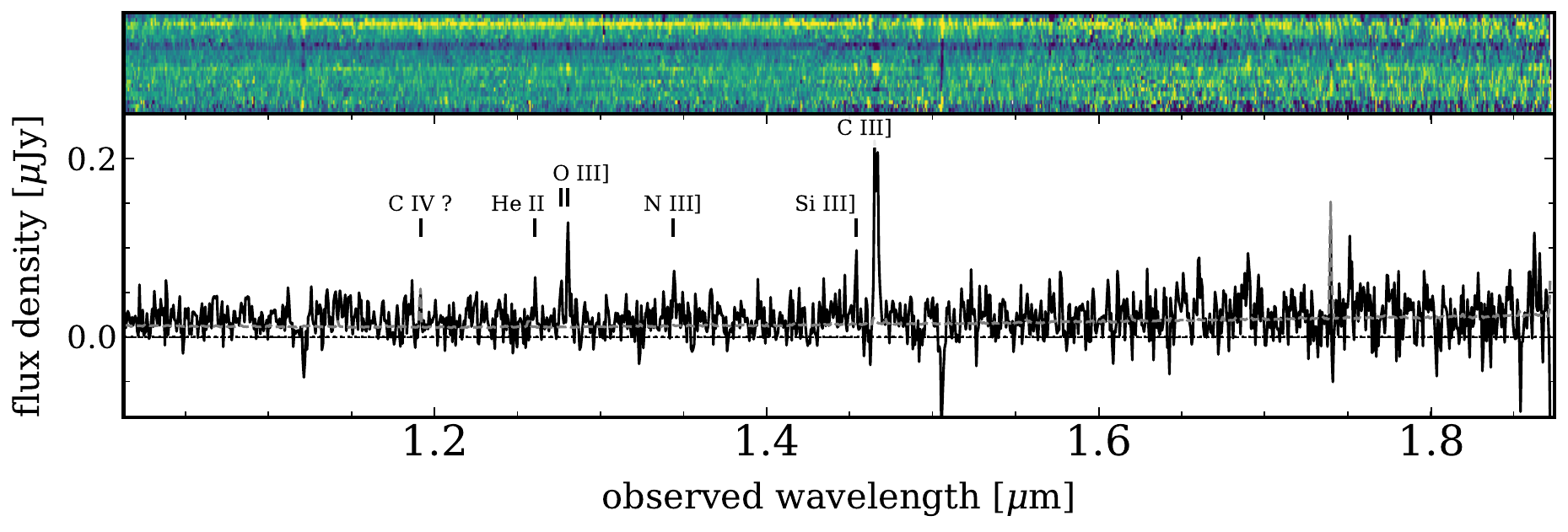}
   \includegraphics[width=0.9\textwidth]{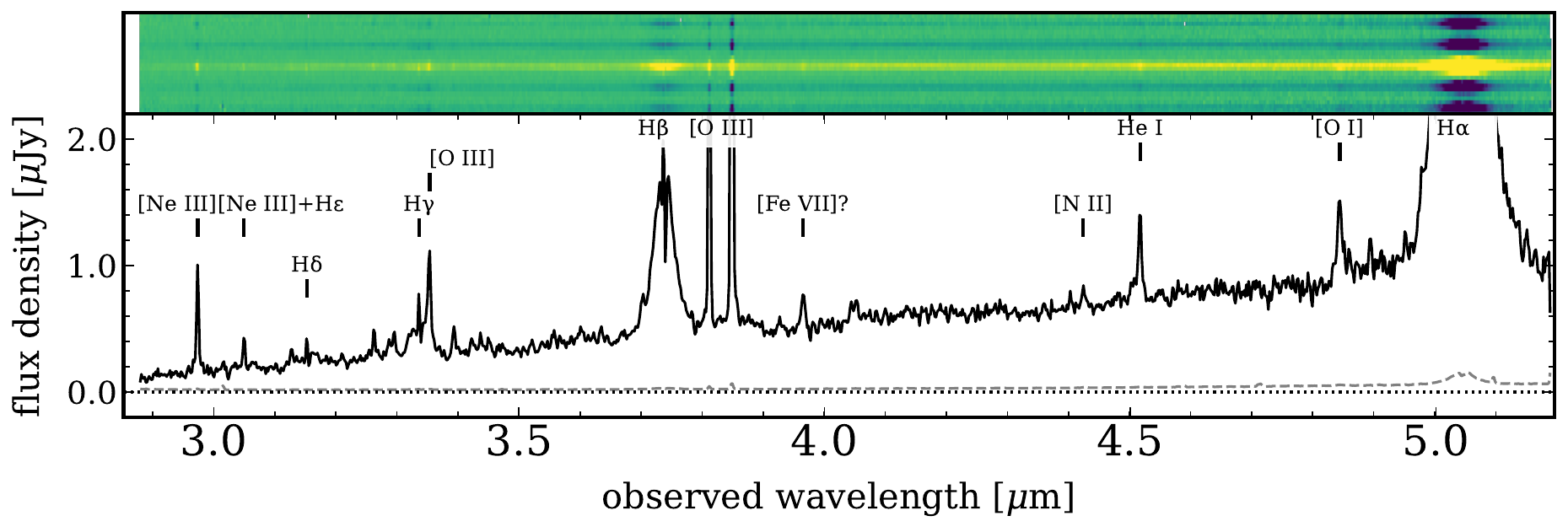}

\end{center}
    \caption{NIRSpec G140M (top panels) and G395M (bottom panels) spectra of C3PO 46403 at $z=6.68$.  The top portion of each panel shows the 2D spectrum, while in the bottom portion, the solid line shows the extracted 1D spectrum, the dashed line shows the uncertainty, and the dotted line shows zero flux density.  The G395M spectrum has an exposure time of 10.3 hrs and covers important rest-optical diagnostic lines, as indicated.   The G140M spectrum has an exposure time of 13.3 hrs and covers important far-UV diagnostic lines.  The spectrum indicates well-detected lines, [\ciii\ $\lambda$1907 + \ciii\ $\lambda$1909, \oiiiuv\ $\lambda\lambda$1661,1666, and \ion{Si}{3} $\lambda$1892, all with S/N $>$3, and less-well detected lines such as \heii\ $\lambda$1640 and \ion{N}{3}] $\lambda\lambda$1746,1748 with $2 < \mathrm{SNR} < 3$. }\label{fig:46403_spec}
\end{figure*}

A second problem is that the broad-emission-line components of LRDs also challenge dense-gas models. LRDs have shown broad emission-line components of of hydrogen, neutral helium, and \ion{O}{1} \citep{Labbe_2024,Kokorev_2026,deGraaff_2026,Lin_2026}.  The broad hydrogen and helium lines are thought to either trace virial motions or gas outflows in LRDs \citep{Matthee_2024,Matthee_2026,Brazzini_2025}.     The interpretation of the broad \ion{O}{1} emission is that it is produced in the same region as H$\alpha$, either resulting from Lyman-$\beta$ fluorescence \citep{deGraaff_2026} or electron-charge exchange \citep{Kokorev_2026}. Interestingly, this implies the gas cocoon of the LRDs are enriched by the products of nucleosynthesis, which disfavors models where they are the direct collapse of primordial clouds into a black hole \citep{Pacucci_2026}.  

Moreover, some LRDs and low-redshift analogs of LRDs show cooler gas under the influence of the LRD central source \citep{Lin_2026}.  Several LRDs illustrate this via emission from iron, particularly [\ion{Fe}{2}], as evidenced in both low-resolution NIRspec/prism data \citep[e.g.,][]{Kokorev_2026,Taylor_2025,Tripodi_2025} and higher-resolution data \citep{Lambrides_2025,DEugenio_2026}.  Interesting, the [\ion{Fe}{2}] lines have velocity widths intermediate of the broad lines and narrow lines in LRDs, implying they originate from a cool, dense gas layer outside the LRD \citep{Lin_2026}.  The existence of forbidden \ion{Fe}{2} differs strongly from detections of permitted \ion{Fe}{2} in quasars \citep{Lin_2026}, and may indicate a fundamental difference between the standard model of AGN and LRDs.  This may hint at the differences in the ``classical'' AGN model and the conditions in LRDs.  

\begin{figure*}
\begin{center}
    \includegraphics[width=0.9\textwidth]{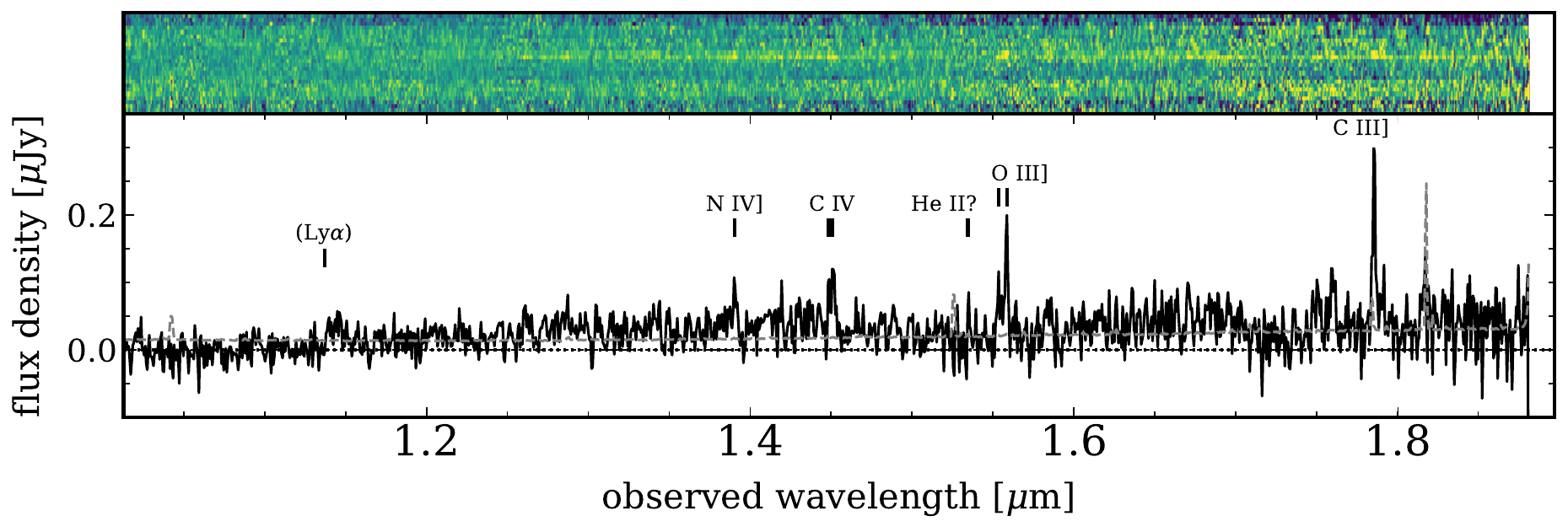}
    \includegraphics[width=0.9\textwidth]{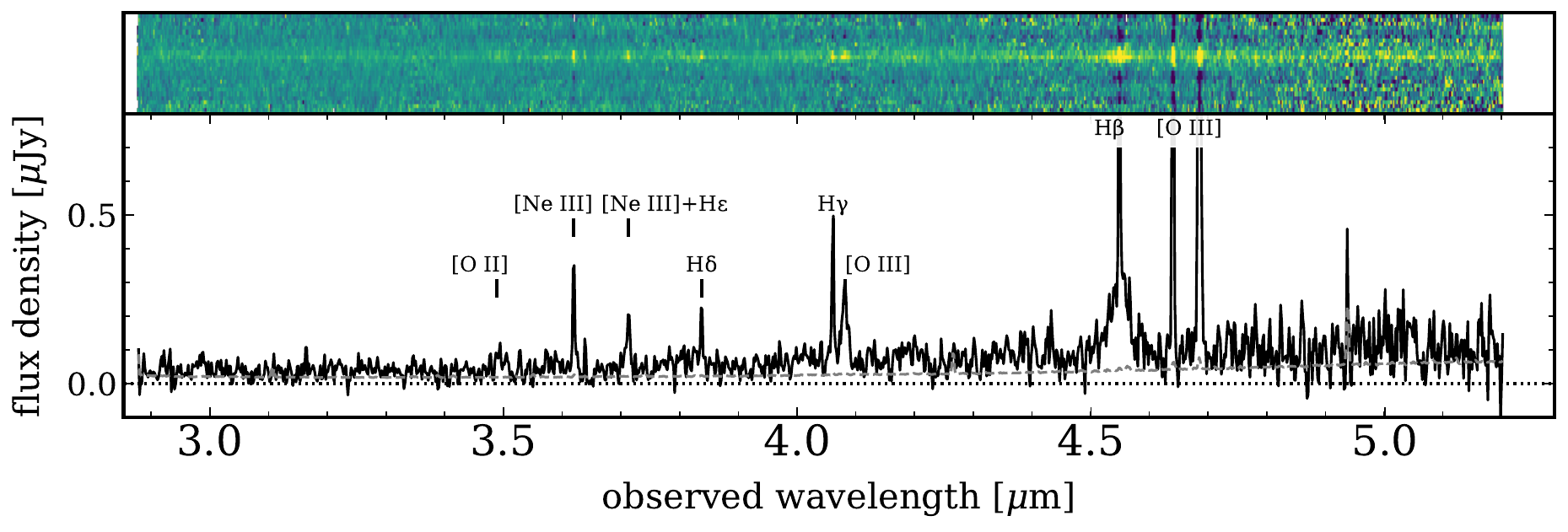}
\end{center}
    \caption{Same as Figure~\ref{fig:46403_spec}, but for C3PO 45290 at $z=8.35$. The exposure times are identical.  Again, important emission lines are indicated.  The G140M spectrum indicates well-detected lines, \civ\ $\lambda\lambda$1548,1551, [\ciii\ $\lambda$1907 + \ciii\ $\lambda$1909, \ion{N}{4}] $\lambda$1486, and \oiiiuv\ $\lambda\lambda$1661,1666, all with S/N $>$3, and \heii\ $\lambda$1640, detected at SNR=2.1.  }\label{fig:45290_spec}
\end{figure*}

Here, we report the analysis of deep, $>$10~hr rest-UV and optical spectra from \jwst/NIRSpec of two LRDs observed by the Carbon-3 plus Oxygen (C3PO) survey.  The LRDs in this sample are C3PO 45290 at $z=8.35$ and C3PO 46403 at $z=6.68$.  C3PO 45290 is one of only three LRDs spectroscopically confirmed at $z > 8$ \citep[cf.][]{Wang_2026,Taylor_2025,Tripodi_2025}, and previously studied using NIRSpec/prism data \citep{Wang_2024,Wang_2026}.    C3PO 46403 is a very bright, well studied LRD (e.g., the ``GlimmIr'', \citealt{Lambrides_2026}, also known as ``Irony'', \citealt{DEugenio_2026}).   While both LRDs show evidence of broad \oiii\ $\lambda$4364 lines in our dataset, many of their other properties differ.  C3PO 46403 is substantially brighter, $m_\mathrm{AB}(F356W) = 24.4$~mag versus $m_\mathrm{AB}(F356W) = 26.9$ for C3PO 45290.  C3PO 46403 shows extreme iron line emission, including [\ion{Fe}{2}] and possible coronal [\ion{Fe}{6}] $\lambda$5176, neither of which is detected in the latter.  Therefore our study of these two objects allows us to compare and contrast the structure of two different LRDs in the early universe.  

The outline for the rest of this \textit{Paper} is as follows.  In Section~\ref{section:data} we describe the C3PO dataset and the observations of the two LRDs here.  In Section~\ref{section:analysis} we discuss the measurements of emission-lines from the G140M and G395M data.   In Section~\ref{section:results} we describe the constraints the measurements place on gas densities.  In Section~\ref{section:discussion} we discuss the implications our measurements have on the ionizing source in LRDs, and the structure of the LRD itself.  In Section~\ref{section:summary}, we summarize our findings.  Throughout we use a flat cosmology with $\Omega_{m,0}=0.315$ and $H_0 = 67.4$~km s$^{-1}$ Mpc$^{-1}$ \citep{Planck_2020}.  All emission-line wavelengths are reported in vacuum with values from NIST \citep{NIST_ASD}.      
%

\section{Data}\label{section:data}

The data used in this work come from the C3PO survey, which is a \jwst/Cycle 3 program (PID: 5943, PIs: Papovich, Hutchison, Hu).  The survey strategy, objectives, and data reduction are described in Papovich et al.\ (in prep).   Briefly, C3PO provides deep NIRSpec MSA observations in G140M (13.3~hrs) and G395M (10.3~hrs) for the highest priority targets to probe the rest-frame UV and optical for galaxies near and into the epoch of reionization (EoR), $7 < z < 9$.  



Two of the primary C3PO targets were selected as LRDs \citep{Kocevski_2025}: C3PO 46403 at $z=6.68$ ($\alpha_{2000} = 14^\mathrm{h}19^\mathrm{m}34.14^\mathrm{s}$, $\delta_{2000} = 52^\circ52^\prime38.7^{\prime\prime}$)  and C3PO 45290 at $z=8.35$ ($\alpha_{2000} = 14^\mathrm{h}19^\mathrm{m}30.27^\mathrm{s}$, $\delta_{2000} = 52^\circ52^\prime51.0^{\prime\prime}$). 
These were the only two LRDs targeted by C3PO that cover \hgamma\ and \oiii\ $\lambda$4364.   As mentioned in Section~\ref{section:introduction}, C3PO 46403 has been studied using NIRSpec prism data and shallower G395M data \citep{Kocevski_2025,Lambrides_2025,Tang_2025,DEugenio_2026,Ronayne_2026}, and the C3PO data for 46403 have been used previously to study variability in this source \citep{Lambrides_2026}.  C3PO 45490 has been studied using NIRSpec/prism data \citep{Wang_2024,Wang_2026}.   
 The C3PO data for both LRDs provide 13.3 hrs in G140M and 10.3 hrs in G395M, achieving the best sensitivities for fainter lines and features in the rest-UV and optical spectra of LRDs to date. 


\begin{deluxetable*}{clr|r|r}
\tablecolumns{5}
\tablewidth{0pt}
\tablecaption{Measured emission line parameters from the G395M spectra for C3PO LRDs used in this work \label{table:optlines}}
\tablehead{ \colhead{} & 
\multicolumn{1}{l}{Parameter} & \multicolumn{1}{r}{Unit} & \colhead{C3PO 45290} & \colhead{C3PO 46403}
}
\startdata
\multirow{11}{*}{\rotatebox{90}{narrow lines}} & redshift\tablenotemark{$\ast$}                & [---]                & 8.3536 $\pm$ 0.0003 & 6.68434 $\pm$ 0.00002 \\
& FWHM$_n$([\ion{O}{3}] $\lambda$5008)  & [km s$^{-1}$] & 235 $\pm$ 3 & 285 $\pm$ 2\\ 
& F$_n$([\ion{O}{2}] $\lambda$3727) + F$_n$([\ion{O}{2}] $\lambda$3730) & {[}$10^{-18}~\mathrm{erg~s^{-1}~cm^{-2}}${]} & 0.054 $\pm$  0.034  & \nodata \\
& F$_n$([\ion{Ne}{3}] $\lambda$3870) & [$10^{-18}~\mathrm{erg~s^{-1}~cm^{-2}}$]  & 0.332 $\pm$ 0.019  &  1.226 $\pm$  0.040 \\
 & F$_n$(H$\gamma$)         & {[}$10^{-18}~\mathrm{erg~s^{-1}~cm^{-2}}${]} & 0.320 $\pm$ 0.020   & 0.347 $\pm$ 0.037 \\
 & F$_n$([\ion{O}{3}] $\lambda$4364) & {[}$10^{-18}~\mathrm{erg~s^{-1}~cm^{-2}}${]} & 0.037 $\pm$ 0.034     & 0.650 $\pm$ 0.061 \\
 & F$_n$(\ion{He}{2} $\lambda$4687) & [$10^{-18}~\mathrm{erg~s^{-1}~cm^{-2}}$] & 0.075 $\pm$ 0.065 & 0.098 $\pm$ 0.024 \\
 & EW$_0$(\ion{He}{2}) & [\AA] & $<$ 11.6 & $<$ 2.4 \\ 
 & F$_n$(H$\beta$)          & {[}$10^{-18}~\mathrm{erg~s^{-1}~cm^{-2}}${]} & 0.482 $\pm$ 0.025   & 0.963 $\pm$ 0.380         \\
 & F$_n$([\ion{O}{3}] $\lambda$4960) & {[}$10^{-18}~\mathrm{erg~s^{-1}~cm^{-2}}${]} & 0.787 $\pm$ 0.031    & 3.107 $\pm$ 0.057 \\
 & F$_n$([\ion{O}{3}] $\lambda$5008) & {[}$10^{-18}~\mathrm{erg~s^{-1}~cm^{-2}}${]} & 2.459 $\pm$ 0.046    & 8.088 $\pm$ 0.087   \\\hline
 \multirow{11}{*}{\rotatebox{90}{broad lines}}& velocity offset, $\Delta$v(H$\beta$)\tablenotemark{$\dag$}  & [km s$^{-1}$] & 65 $\pm$ 107 &  84 $\pm$ 180 \\ 
 & velocity offset, $\Delta$v(\oiii\ $\lambda$4364)\tablenotemark{$\dag$} & [km s$^{-1}$] & $-106$ $\pm$ 67 & \nodata \\ 
 & velocity offset, $\Delta$v([\ion{Fe}{2}])\tablenotemark{$\dag$}  & [km s$^{-1}$] & \nodata & 43 $\pm$ 75 \\ 
 & FWHM$_b$(H$\beta$)          & [km s$^{-1}$] & 2549 $\pm$ 216 & 3110 $\pm$ 56 \\ 
 & FWHM$_b$([\ion{O}{3}] $\lambda$4364) & [km s$^{-1}$] & 870 $\pm$ 139 & 1119 $\pm$ 28 \\ 
 & FWHM$_b$([\ion{Fe}{2}])  & [km s$^{-1}$] & \nodata & 399 $\pm$ 53 \\
 & F$_b$(H$\gamma$)         & {[}$10^{-18}~\mathrm{erg~s^{-1}~cm^{-2}}${]} & 0.031 $\pm$ 0.071     &  2.210 $\pm$ 0.272 \\
 & F$_b$([\ion{O}{3}] $\lambda$4364) & {[}$10^{-18}~\mathrm{erg~s^{-1}~cm^{-2}}${]} & 0.396 $\pm$ 0.068     &  0.511 $\pm$ 0.185 \\
 & F$_b$(H$\beta$)          & {[}$10^{-18}~\mathrm{erg~s^{-1}~cm^{-2}}${]} & 1.227 $\pm$ 0.142     &  11.348 $\pm$ 0.292 \\
 & F$_b$([\ion{O}{3}] $\lambda$4960)  & {[}$10^{-18}~\mathrm{erg~s^{-1}~cm^{-2}}${]} & 0.110 $\pm$ 0.073     &  0.319 $\pm$ 0.057 \\
 & F$_b$([\ion{O}{3}] $\lambda$5008)  & {[}$10^{-18}~\mathrm{erg~s^{-1}~cm^{-2}}${]} & 0.326 $\pm$ 0.073     &  0.949 $\pm$ 0.294 \\\hline
\enddata
\tablenotetext{\ast}{Computed from the \oiii\ $\lambda\lambda$4960, 5008 narrow lines.}
\tablenotetext{$\dag$}{Computed relative to the \oiii\ $\lambda\lambda$4960, 5008 narrow lines.}
\end{deluxetable*}

\begin{deluxetable*}{lr|r|r}
\tablecolumns{5}
\tablewidth{0pt}
\tablecaption{Measured Emission Line Parameters from the G140M spectra for C3PO LRDs\label{table:uvlines}}
\tablehead{ 
\multicolumn{1}{l}{Parameter} & \multicolumn{1}{r}{Unit} & \colhead{C3PO 45290} & \colhead{C3PO 46403}
}
\startdata
redshift\tablenotemark{$\ast$}                & [---]                & 8.3532 $\pm$ 0.0013 &  6.6847 $\pm$ 0.0014 \\
 velocity offset, $\Delta$v(\civ)\tablenotemark{$\dag$} & [km s$^{-1}$] & $111$ $\pm$ 96 & 437 $\pm$ 223 \\ 
 velocity offset, $\Delta$v(\heii)\tablenotemark{$\dag$}  & [km s$^{-1}$] & 118 $\pm$ 81 & 30 $\pm$ 104 \\ 
F(\ion{N}{4} $\lambda$1483)         & [$10^{-19}~\mathrm{erg~s^{-1}~cm^{-2}}$] & 0.62 $\pm$ 0.41 & \nodata \\
F(\ion{N}{4} $\lambda$1486)         & [$10^{-19}~\mathrm{erg~s^{-1}~cm^{-2}}$] & 1.81 $\pm$ 0.41 & \nodata \\
F(\civ\ $\lambda$1548)         & [$10^{-19}~\mathrm{erg~s^{-1}~cm^{-2}}$] & 1.48 $\pm$ 0.42 & 0.69 $\pm$ 0.45\tablenotemark{$\ddag$} \\
F(\civ\ $\lambda$1551)         & [$10^{-19}~\mathrm{erg~s^{-1}~cm^{-2}}$] & 2.98 $\pm$ 0.65 & \nodata  \\
F(\heii\ $\lambda$1640)         & [$10^{-19}~\mathrm{erg~s^{-1}~cm^{-2}}$] & 1.01 $\pm$ 0.48 & 1.02 $\pm$ 0.51 \\
F(\oiiiuv\ $\lambda$1661)         & [$10^{-19}~\mathrm{erg~s^{-1}~cm^{-2}}$] & 1.44 $\pm$ 0.36 & 1.28 $\pm$ 0.35 \\
F(\oiiiuv\ $\lambda$1666)         & [$10^{-19}~\mathrm{erg~s^{-1}~cm^{-2}}$] & 3.43 $\pm$ 0.36 & 3.02 $\pm$ 0.35 \\
F(\ion{N}{3}] $\lambda\lambda$1746,1748)  & [$10^{-19}~\mathrm{erg~s^{-1}~cm^{-2}}$] & \nodata  & 1.52 $\pm$ 0.60 \\
F(\ion{Si}{3}] $\lambda$1883)         & [$10^{-19}~\mathrm{erg~s^{-1}~cm^{-2}}$] & \nodata & 0.63 $\pm$ 0.46 \\
F(\ion{Si}{3}] $\lambda$1892)         & [$10^{-19}~\mathrm{erg~s^{-1}~cm^{-2}}$] & \nodata & 1.40 $\pm$ 0.47 \\
F([\ion{C}{3}] $\lambda$1907)         & [$10^{-19}~\mathrm{erg~s^{-1}~cm^{-2}}$] & 1.34 $\pm$ 0.77 & 3.56 $\pm$ 0.63 \\
F(\ciii\ $\lambda$1909)               & [$10^{-19}~\mathrm{erg~s^{-1}~cm^{-2}}$] & 3.91 $\pm$ 1.77 & 3.76 $\pm$ 0.63 \\
EW$_0$(\ion{N}{4}] $\lambda$1486)  & [\AA] & 3.8 $\pm$ 0.5 &  \nodata  \\
EW$_0$(\civ\ $\lambda\lambda$1548,1551)  & [\AA] & 12.0 $\pm$ 0.6 & 0.6 $\pm$ 1.0  \\
EW$_0$(\heii\ $\lambda$1640)  & [\AA] & 2.4 $\pm$ 0.6 & 5.9 $\pm$ 0.4  \\
EW$_0$(\oiiiuv\ $\lambda$1661)  & [\AA] & 8.5 $\pm$ 0.5 & 7.2 $\pm$ 0.5  \\
EW$_0$(\oiiiuv\ $\lambda$1666)  & [\AA] & 16.0 $\pm$ 0.5 & 14.7 $\pm$ 0.4  \\
EW$_0$(\ion{N}{3}] $\lambda\lambda$1746,1748) &  [\AA]  & \nodata & 9.3 $\pm$ 0.5 \\ 
EW$_0$(\ion{Si}{3}] $\lambda$1883) & [\AA] & \nodata & 3.6 $\pm$ 0.4 \\ 
EW$_0$(\ion{Si}{3}] $\lambda$1892) & [\AA] & \nodata & 5.6 $\pm$ 0.4 \\ 
EW$_0$([\ion{C}{3}] $\lambda$1907 + \ciii\ $\lambda$1909) & [\AA] & 19.2 $\pm$ 0.7 & 40.4 $\pm$ 0.5 \\\hline
\enddata
\tablenotetext{\ast}{Computed from the \oiiiuv\ $\lambda\lambda$1661, 1666 lines.}
\tablenotetext{$\dag$}{Computed relative to the \oiiiuv\ $\lambda\lambda$1661, 1666 lines.  Other lines show no statistically significant offsets.}
\tablenotetext{$\ddag$}{Flux reported for the sum of the \civ\ $\lambda$1548 + \civ\ $\lambda$1551 lines in C3PO 46403.}
\end{deluxetable*}

The data reduction and spectral extraction steps are described elsewhere (Hu et al., in prep., Papovich et al., in prep.).   The general processing is based on standard pipeline with several modifications (see \citealt{Arrabal_Haro_2023a,Hutchison_2026}).   Briefly, we reduced the data using the standard \jwst\ pipeline \citep[version 1.20.2;][]{Bushouse_2025} with the Calibration Reference Data System (CRDS) version \texttt{jwst\_1464.pmap}.   For the two LRDs here, we applied an optimal extraction location for each object by collapsing the 2D spectrum in the spectral direction and fitting a Gaussian to the light profile in the spatial direction at the expected location of the target.  We applied the default pipeline pathloss  correction.  This assumes that the targets are point sources, which is appropriate for the LRDs here.  

Figures~\ref{fig:46403_spec} and \ref{fig:45290_spec} show the 2D and 1D spectra from G140M and G395M for each LRD, C3PO 46403 and C3PO 45290, respectively.   The G395M data show broad Balmer lines and narrow metal lines, consistent with the LRD population.  Both objects show detections of a large suite of emission lines, including many Balmer lines, the \oiii\ $\lambda\lambda$4960,5008 and auroral \oiii\ $\lambda$4364 lines, and [\ion{Ne}{3}] $\lambda$3870 in the G395M grating.  The G140M grating shows many metal lines including [\ciii\ $\lambda$1907, \ciii\ $\lambda$1909, \oiiiuv\ $\lambda\lambda$1661,1666, and \heii\ $\lambda$1640, as well as other diagnostic lines.  Remarkably, in both LRDs, the G395M observations show evidence of broad \oiii\ $\lambda$4364 auroral lines, significantly broader than the \oiii\ $\lambda\lambda$4960,5008 and other metal lines, which show only narrow components.    This observation has important implications for the density of the gas in the broad-line regions of these LRDs. 

\section{Analysis}\label{section:analysis} 

\begin{figure*}[t]
\begin{center}
    \includegraphics[width=0.45\textwidth]{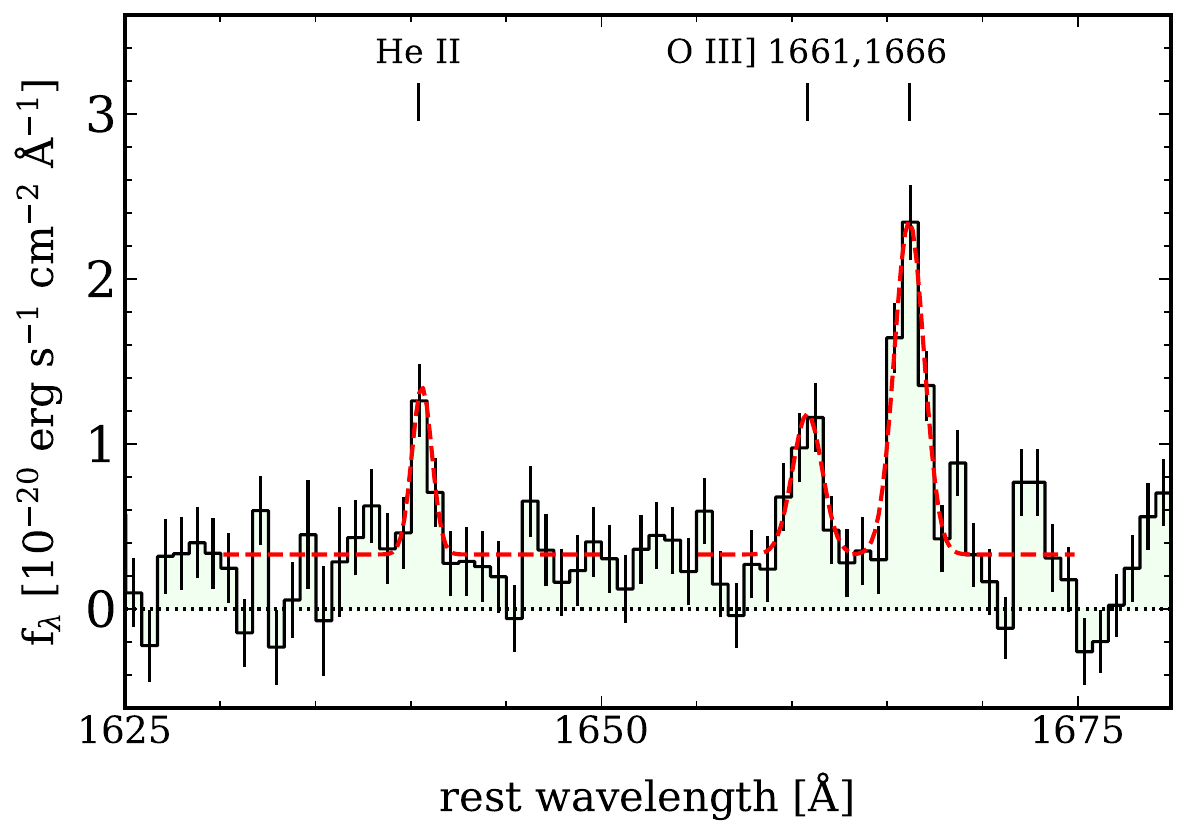}
    \includegraphics[width=0.45\textwidth]{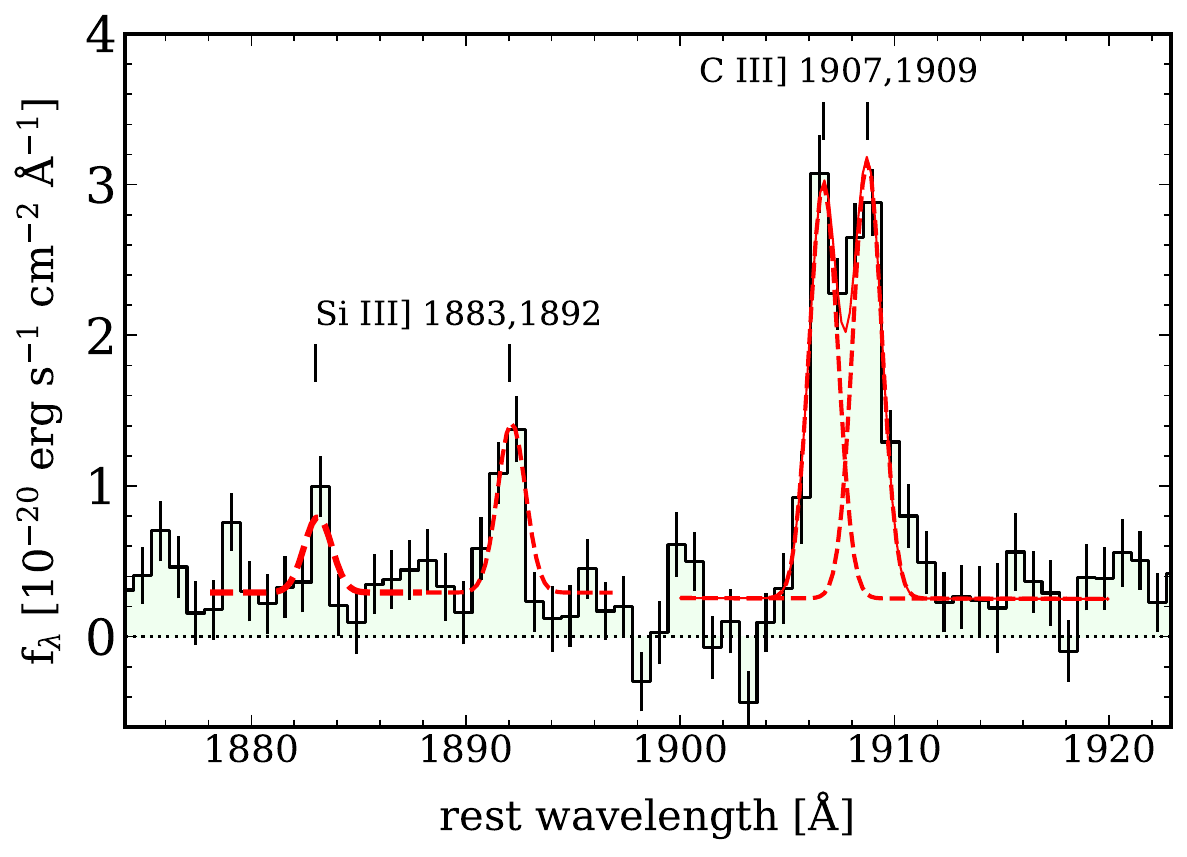}
    \includegraphics[width=0.45\textwidth]{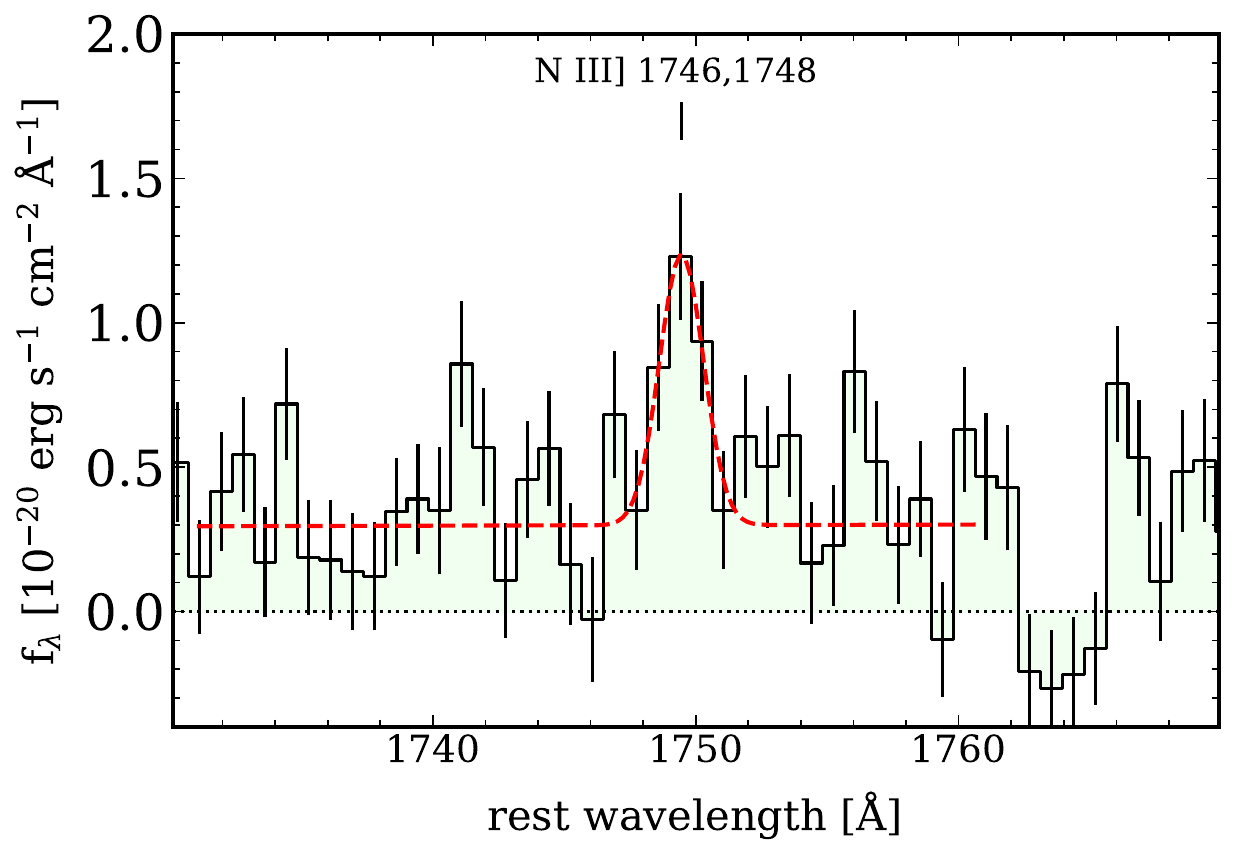}
\end{center}
    \caption{Regions around the detected UV emission lines in the G140M spectrum of C3PO 46403.  The top, left panel shows the region around \heii\ and \oiiiuv, all with significant detections.  The  top, right panel shows the region around \ciii, which shows a strong detection, along with a detection of the \ion{Si}{3}] $\lambda$1892 of the 1883,1892 doublet. The bottom panel shows the region around \ion{N}{3}] $\lambda\lambda$1746,1748.   In each panel, the red curve shows a Gaussian fit to the emission lines and continua.}\label{fig:46403_uvzoom}
\end{figure*}

\begin{figure*}
\begin{center}
    \includegraphics[width=0.9\textwidth]{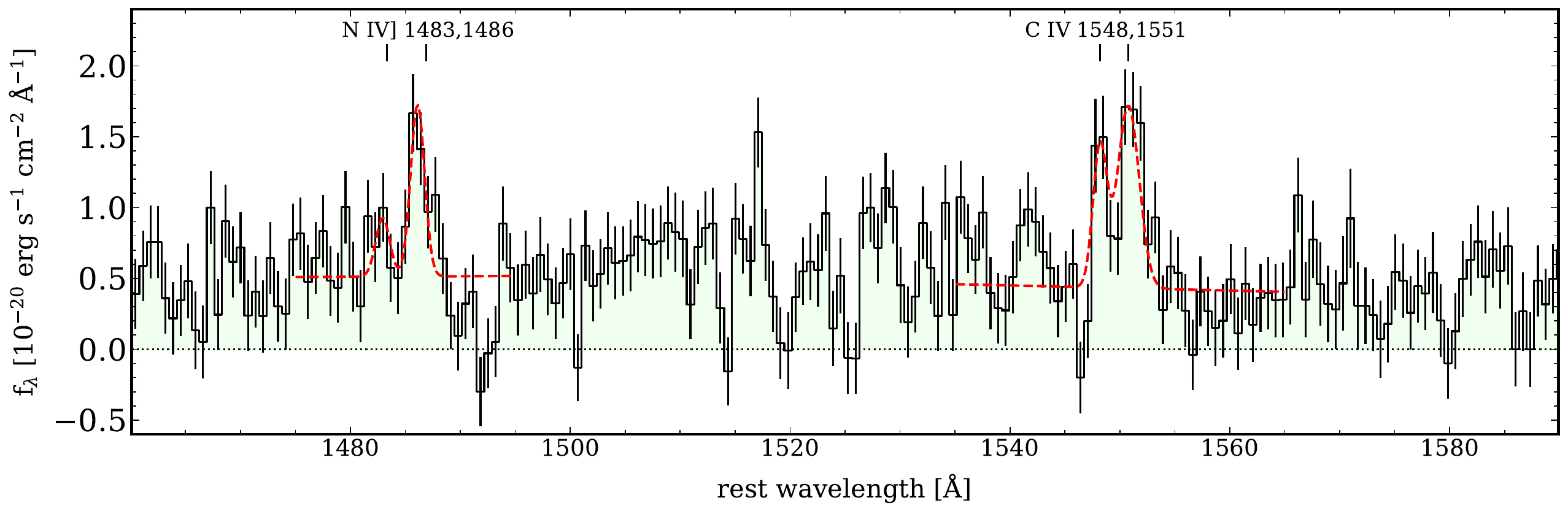}
    \includegraphics[width=0.45\textwidth]{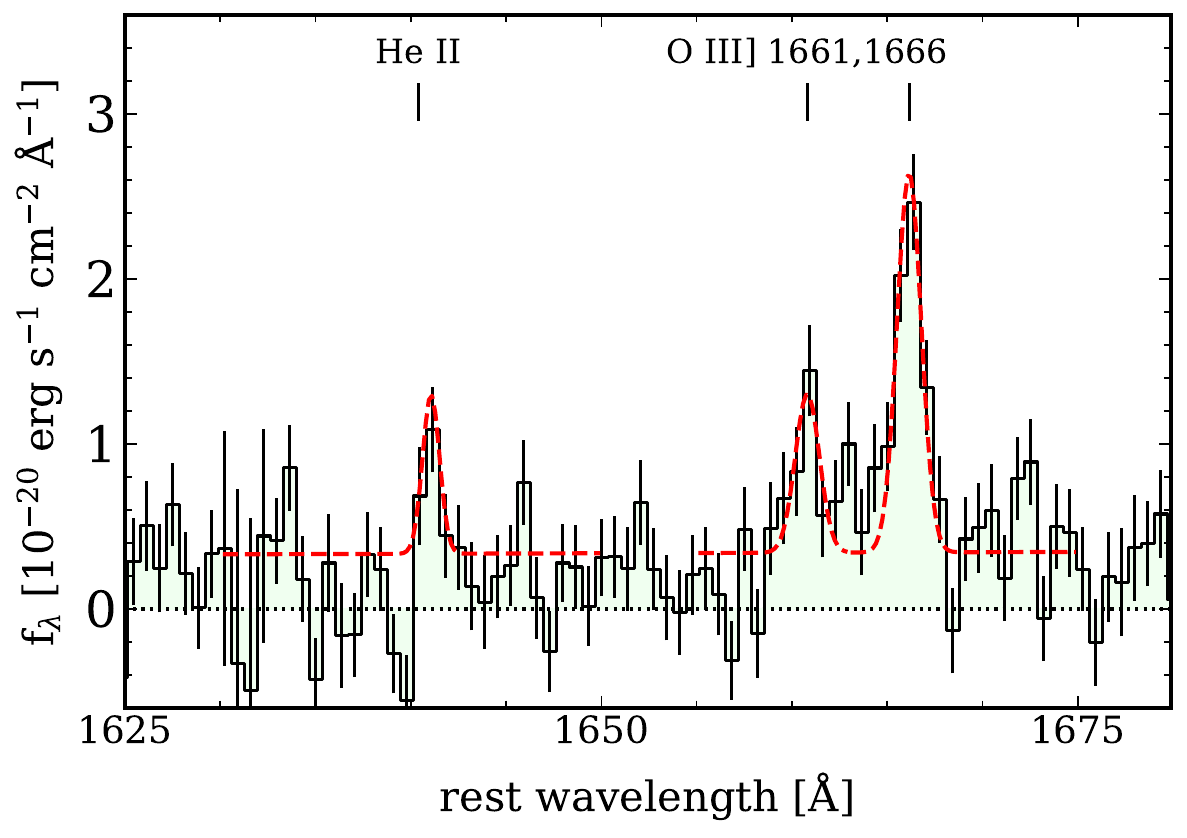}
    \includegraphics[width=0.45\textwidth]{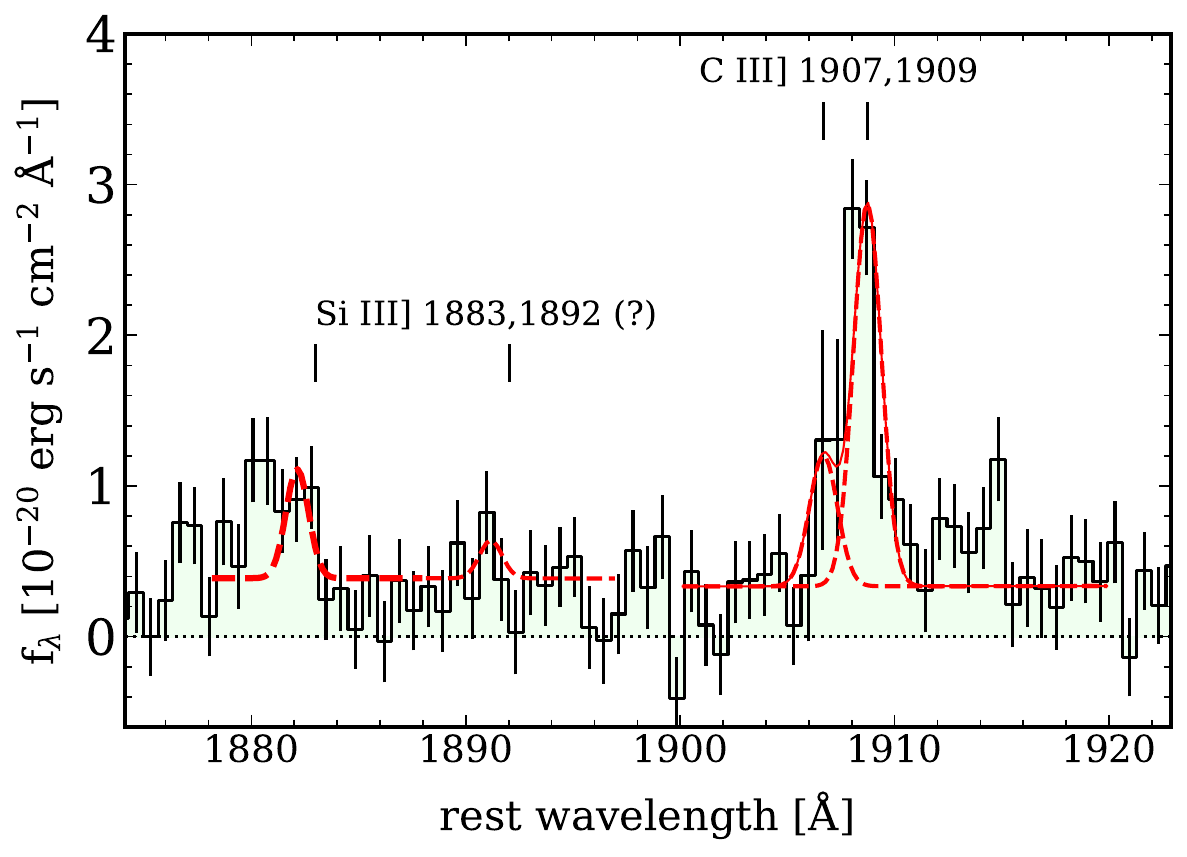}
    \end{center}
    \caption{Regions around the locations of detected UV emission lines in the G140M spectrum of C3PO 45290.  The top panel shows the regions of \ion{N}{4}] and \ion{C}{4}.  The bottom left plot shows the region around \heii\ and \oiiiuv.  The bottom right plot shows the region around \ciii. In each panel, the red curve shows a Gaussian fit to the emission lines and continua.}\label{fig:45290_uvzoom}
\end{figure*}

\begin{figure*}[t]
    \begin{center}
        \includegraphics[width=0.49\textwidth]{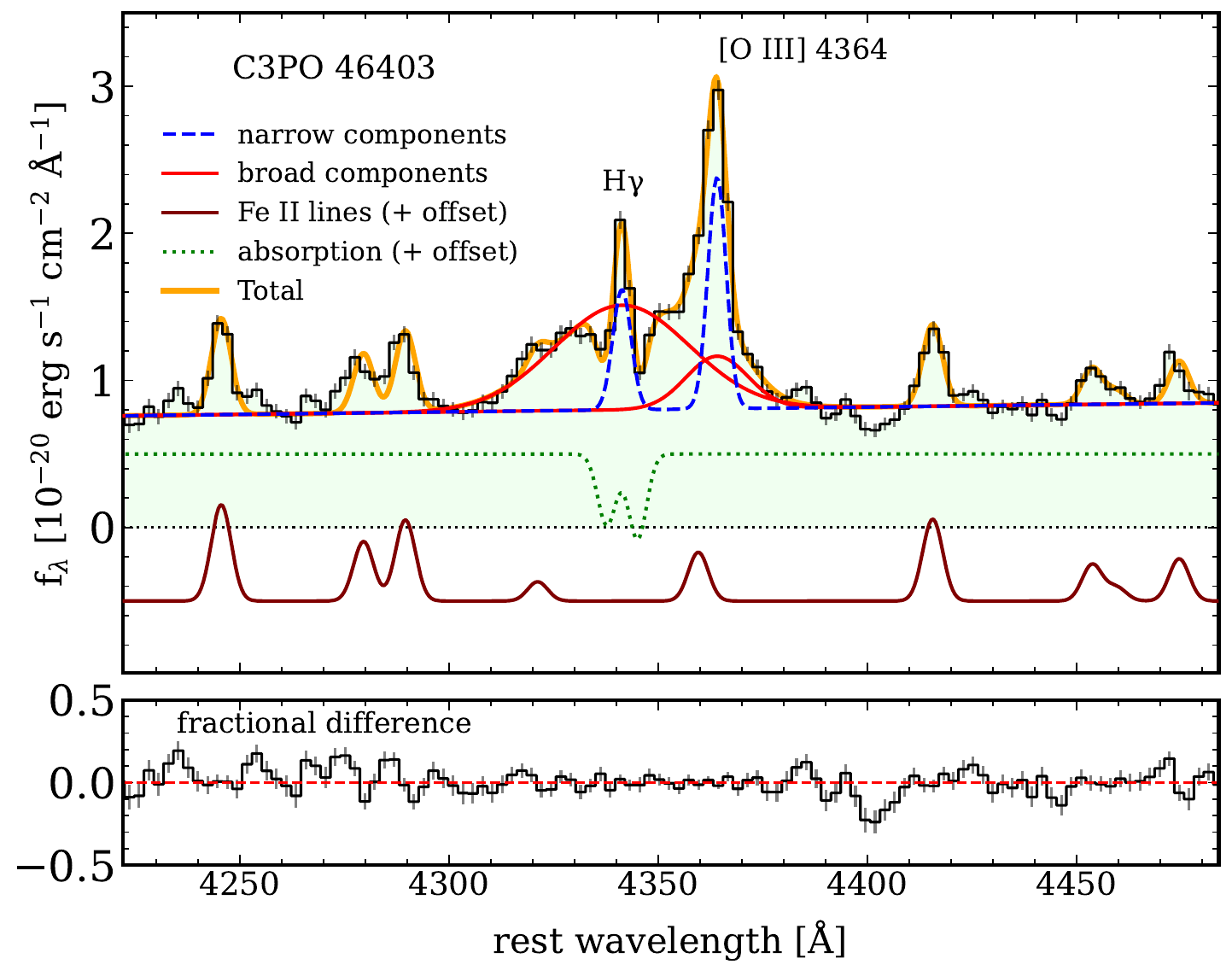}
        \includegraphics[width=0.49\textwidth]{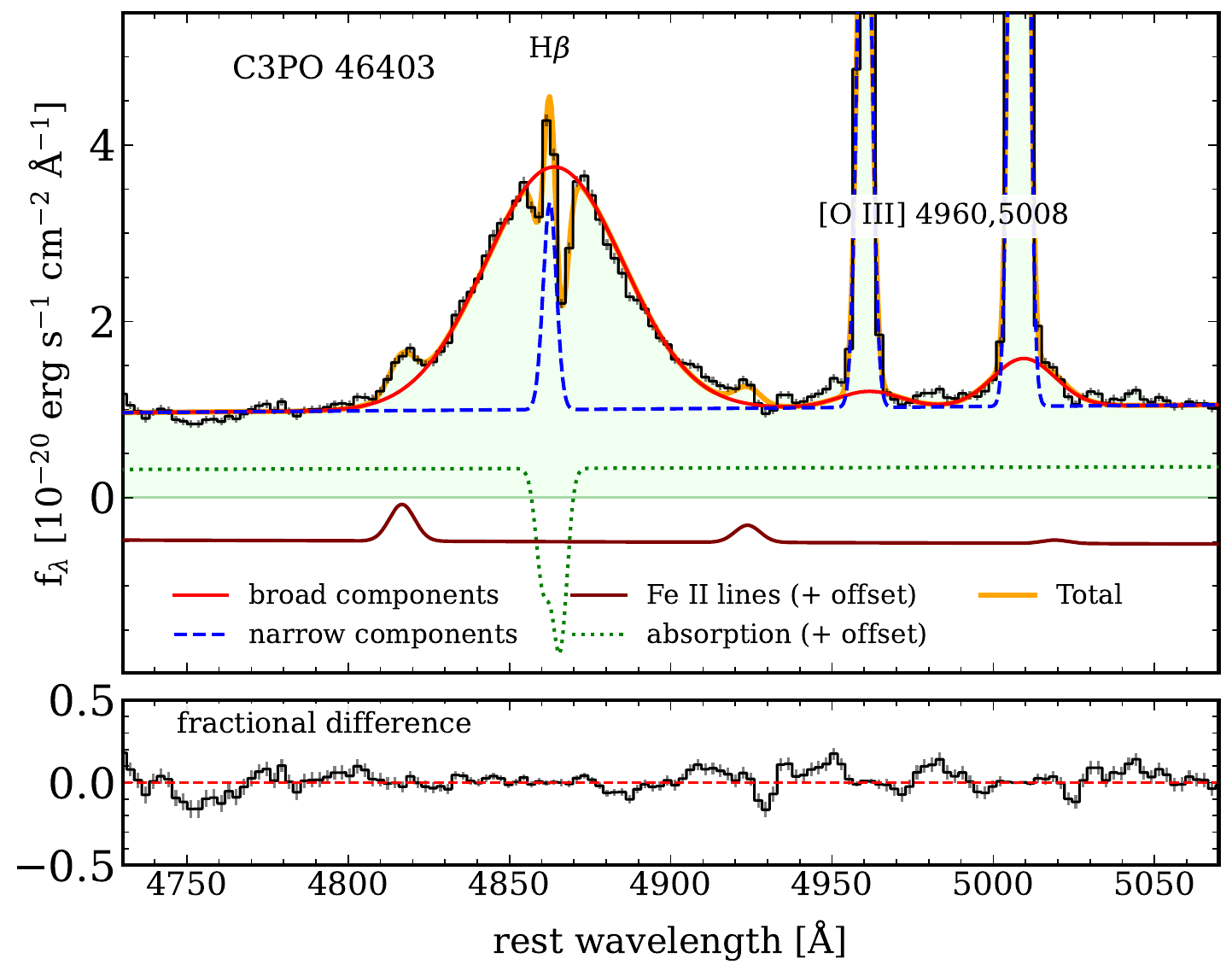}
    \end{center}
    \caption{Spectral region around \hgamma+\oiii\ $\lambda$4364 (left plot) and \hbeta+\oiii\ $\lambda\lambda$4960,5008 (right plot) in the G395M spectrum of C3PO 46403.  In each plot, the black lines and errors show the measured data and uncertainties.  The solid-red curve (dashed-blue curve) shows broad (narrow)-component fits to the lines and continua. The maroon-solid line shows the modeled [\ion{Fe}{2}] lines (offset for clarity).  The green-dotted line shows the Balmer absorption compnonets (offset for clarity).    The thick orange line shows the total model fit to the spectrum.  The bottom panel of each plot shows the fractional residual between the data and the fit.  \hgamma, \hbeta, and \oiii\ 4364 show both broad and narrow components.   }\label{fig:46403_balmerOiii}
\end{figure*}

\begin{figure*}[t]
    \begin{center}
        \includegraphics[width=0.49\textwidth]{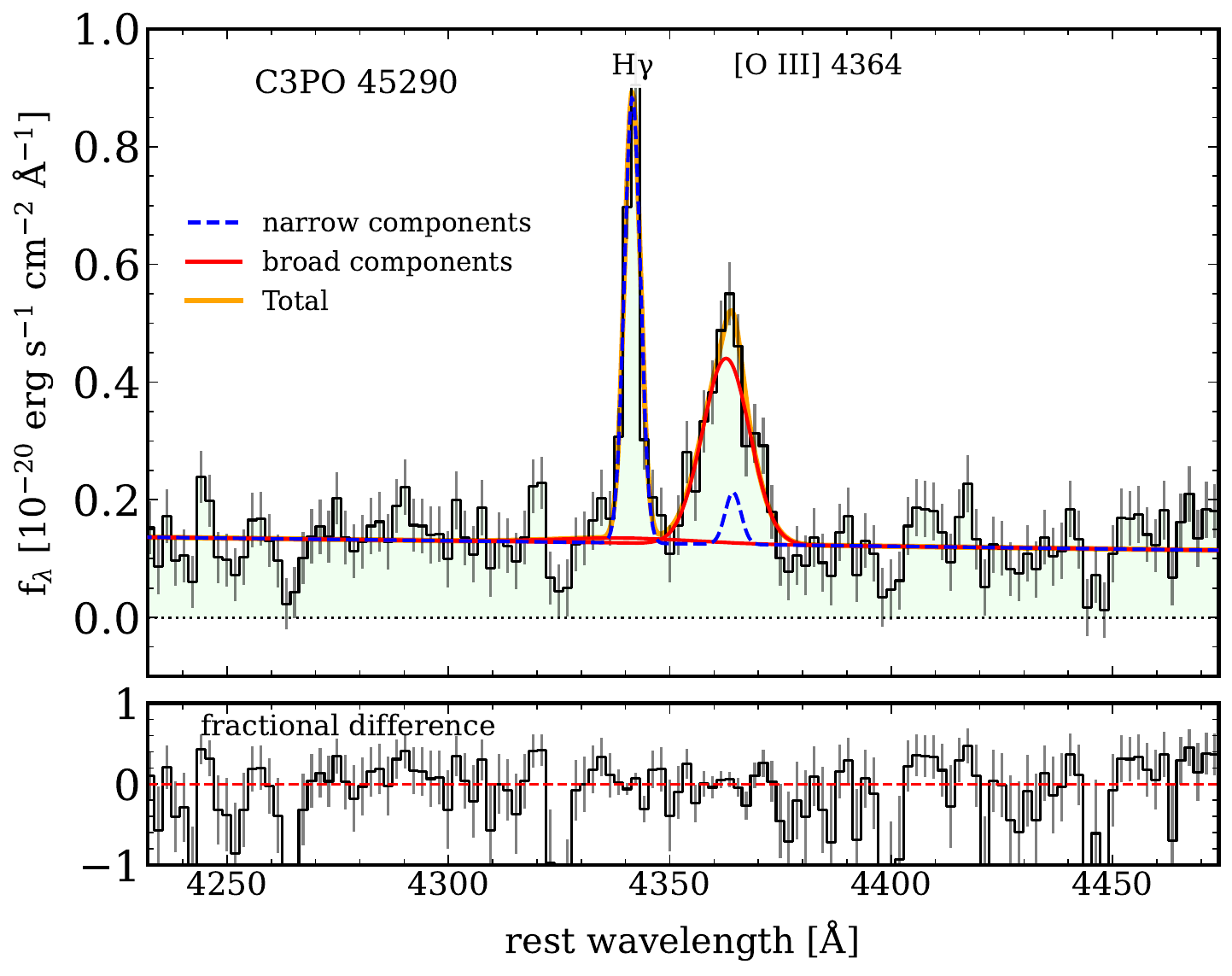}
        \includegraphics[width=0.49\textwidth]{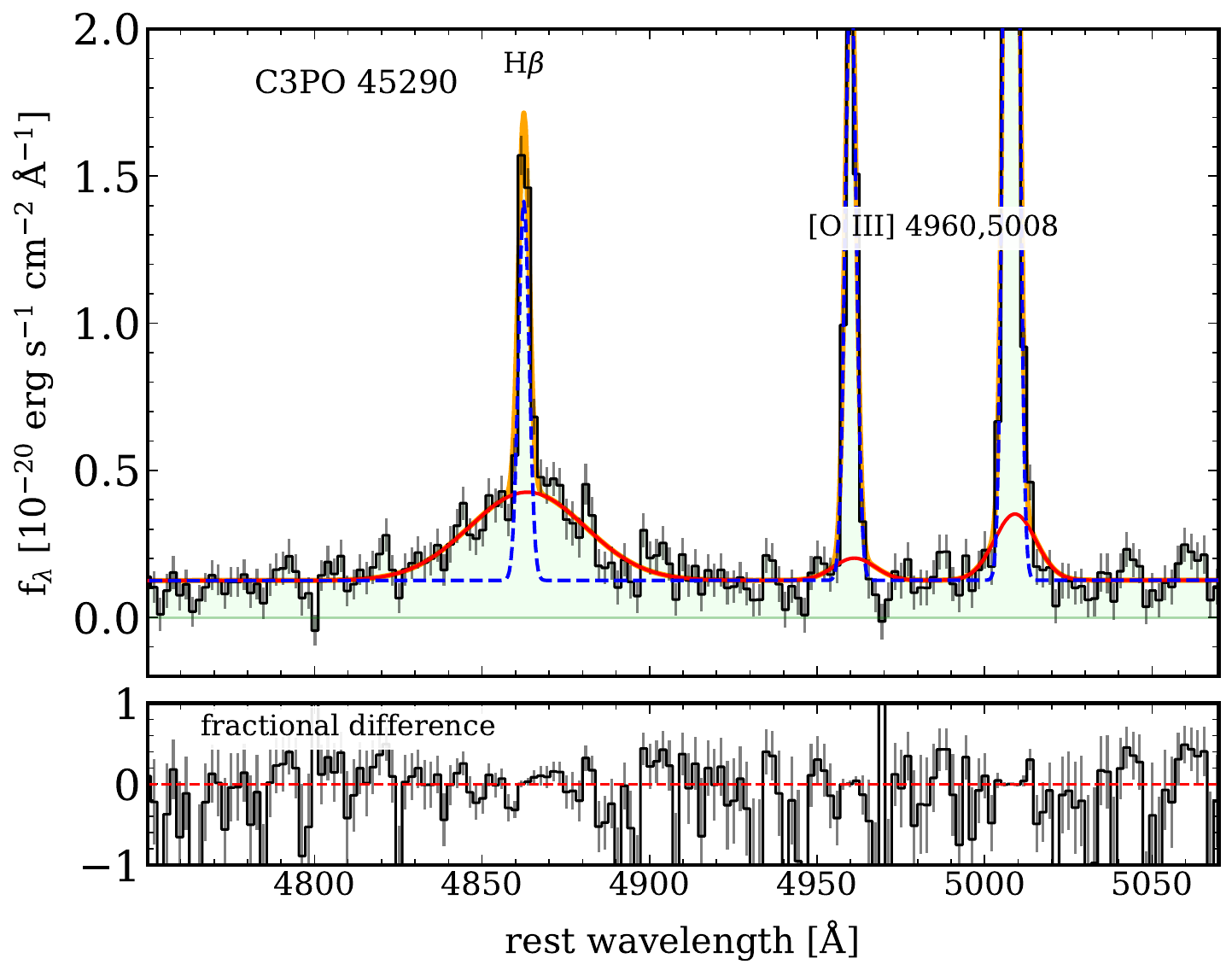}
    \end{center}
    \caption{Same as Figure~\ref{fig:46403_balmerOiii}, but showing the results for C3PO 45290. Unlike C3PO 46403, we do not include any [\ion{Fe}{2}] emission nor Balmer absorption components. }\label{fig:45290_balmerOiii}
\end{figure*}

We fit the emission lines in the spectra, in most cases using a single Gaussian model.  For cases where lines are blended or have lines nearby (such that they would impact the continuum of adjacent lines) we fit them simultaneously.    For the cases of where broad lines are evident (\hbeta, \hgamma, and \oiii\ $\lambda$4364), or to obtain limits (\oiii\ $\lambda\lambda$4960,5008), we model the lines with two Gaussian components to allow for a narrow and broad component, which we fit simultaneously.\footnote{We define ``broad'' and ``narrow'' lines as those having a FWHM greater than or less than $\approx$300 km~s$^{-1}$, respectively, where 300~km~s$^{-1}$ is approximately the spectral resolution of the G140M and G395M gratings. This definition allows us to separate the broader [\ion{Fe}{2}] lines in C3PO 46403 from the narrower rest-UV and optical lines.}  For C3PO 46403, there is evidence of Balmer absorption, which we model by adding additional Gaussian components.  In all cases, we simultaneously fit the continuum using a 1st--order polynomial.   Tables~\ref{table:optlines} and \ref{table:uvlines} provide measured line fluxes, velocity widths, and velocity offsets from systemic for lines used in this work, measured from the G395M and G140M data, respectively. 
Figures~\ref{fig:46403_uvzoom} and \ref{fig:45290_uvzoom} show the regions around detected UV emission lines in C3PO 46403 and 45290, along with the emission-line fits respectively.   Figures~\ref{fig:46403_balmerOiii} and \ref{fig:45290_balmerOiii} show regions and model fits around the \hgamma+\oiii\ $\lambda$4364 and \hbeta+\oiii\ $\lambda\lambda$4960,5008 lines for C3PO 46403 and 45290, respectively.  

The \oiii\ $\lambda$4364 and Balmer lines in both LRDs show evidence of a broad component superimposed on a narrow component. %
%
The broad lines in the LRDs here appear to be well described by a double Gaussian model, as the residuals in the bottom panels of Figure~\ref{fig:46403_balmerOiii} and \ref{fig:45290_balmerOiii} are consistent with noise (with some absorption in the \hbeta\ lines, which our models to not include).   This is consistent with the findings of \cite{Juodzbalis_2025} who find double Gaussian profiles typically fit the broad lines as well or better than exponential models. Furthermore, \citet{DEugenio_2026} show in study of C3PO 46403 using independent data that  fits to the Balmer lines prefer a double Gaussian model compared to an exponential model (though see \citealt{Rusakov_2026}).  We note that this may be a result of our relatively low spectral resolution, $R\sim 1000$, and higher-spectral resolution observations are needed to study better this issue.    Regardless, as the exact detailed shape of the broad lines is less important to our conclusions here,  we adopt the double Gaussian fits to the \oiii\ and Balmer lines and defer a detailed study of their profile to a future work. 

In our analysis we use the line ratios to explore constraints on ionization, temperature, and density in the gas in the narrow-line and broad-line components in these LRDs.  One caveat is that we do not apply a correction for dust attenuation.   Dust in LRDs is a complicated subject, where the interpretation of the Balmer and Paschen line ratios from the broad hydrogen lines is consistent with complex radiative transfer \citep{DEugenio_2026,Lin_2026} where many studies now conclude there is little dust of the LRD source, $A(V) \lesssim 0.5$~mag \citep[e.g.][]{DEugenio_2026,Ronayne_2026}.  The ratio of \hgamma/\hbeta\ in the LRDs here are 0.55 $\pm$ 0.06 and 0.36 $\pm$ 0.15 for C3PO 45290 and 46403, respectively.  These are both consistent with a theoretical value of \hgamma/\hbeta = 0.48 indicating no evidence for significant dust attenuation in the narrow-line regions.\footnote{The ratio of \hgamma/\hbeta\ for C3PO 45290 slightly \textit{exceeds} the theoretical value, which may indicate a more complex situation, or indicate uncertainties in the line modeling.}   The UV spectral slopes, $\beta_\mathrm{UV}$, tell a different story, where we measure $\beta_\mathrm{UV} = -0.8 \pm 0.2$ for both LRDs based on SED fitting of photometric and NIRSpec/prism data \citep{Papovich_2026}.  This implies larger attenuation, $A(V) = 0.7$~mag. However, it is unclear if scattered light from the LRD contributes to the rest-UV continuum, which may redden $\beta_\mathrm{UV}$ without dust \citep{Labbe_2025}.  Therefore, because the Balmer decrements indicate no dust attenuation we do not include any.  However, in our analysis, we consider how our results would change if the dust attenuation is $A(V) = 0.7$~mag, which would have a minor impact the UV line ratios and elemental abundances (see Section~\ref{section:discussion}).

\section{Constraints on LRD Gas Densities}\label{section:results}

\subsection{Narrow-Line Components}\label{section:narrow_gas}

Some of the UV emission lines detected in our data are sensitive to the electron density in different ionization states.  The [\ciii\ $\lambda$1907 and \ciii\ $\lambda$1909 lines originate from two closely spaced fine-structure levels of the same excited configuration, $2s2p$ $^3P$, but decay to the same ground state, $2s^2$ $^1S_0$ via two mechanisms.  The  [\ciii\ $\lambda$1907 line is a strictly forbidden magnetic quadrupole transition, $^3P_2 \rightarrow ^1S_0$, while the \ciii\ $\lambda 1909$ line is a semi-forbidden electric dipole transition $^3P_1 \rightarrow ^1S_0$.  Because the difference in the energy levels of the excited states is small, $\Delta E \approx 0.01$~eV, compared to the typical electron temperature, $E \sim~0.862 (T/10,000~\mathrm{K})$~eV, the collisional excitation rate of the \ciii\ lines is mostly independent of the gas temperature, $T_e$.  Because the critical densities of the lines are $\log n_\mathrm{crit} / \mathrm{cm}^{-3} \approx 5.2$ for [\ion{C}{3}] $\lambda$1907 compared to $\log n_\mathrm{crit} / \mathrm{cm}^{-3} \approx 9.0$ for \ciii\ $\lambda$1909, the ratio of [\ion{C}{3}] $\lambda$1907 / \ciii\ $\lambda$1909 is a strong measure of the gas density.  Both lines are detected in C3PO 46403 and C3PO 45290.   Similar situations exist for other UV line doublets, notably \ion{Si}{3}] $\lambda\lambda$1883, 1892 and \ion{N}{4} $\lambda\lambda$1483, 1486, detected in C3PO 46403 and 45290, respectively. 

\begin{figure}
    \centering
    \includegraphics[width=0.95\linewidth, trim={0 48pt 0 0}, clip]{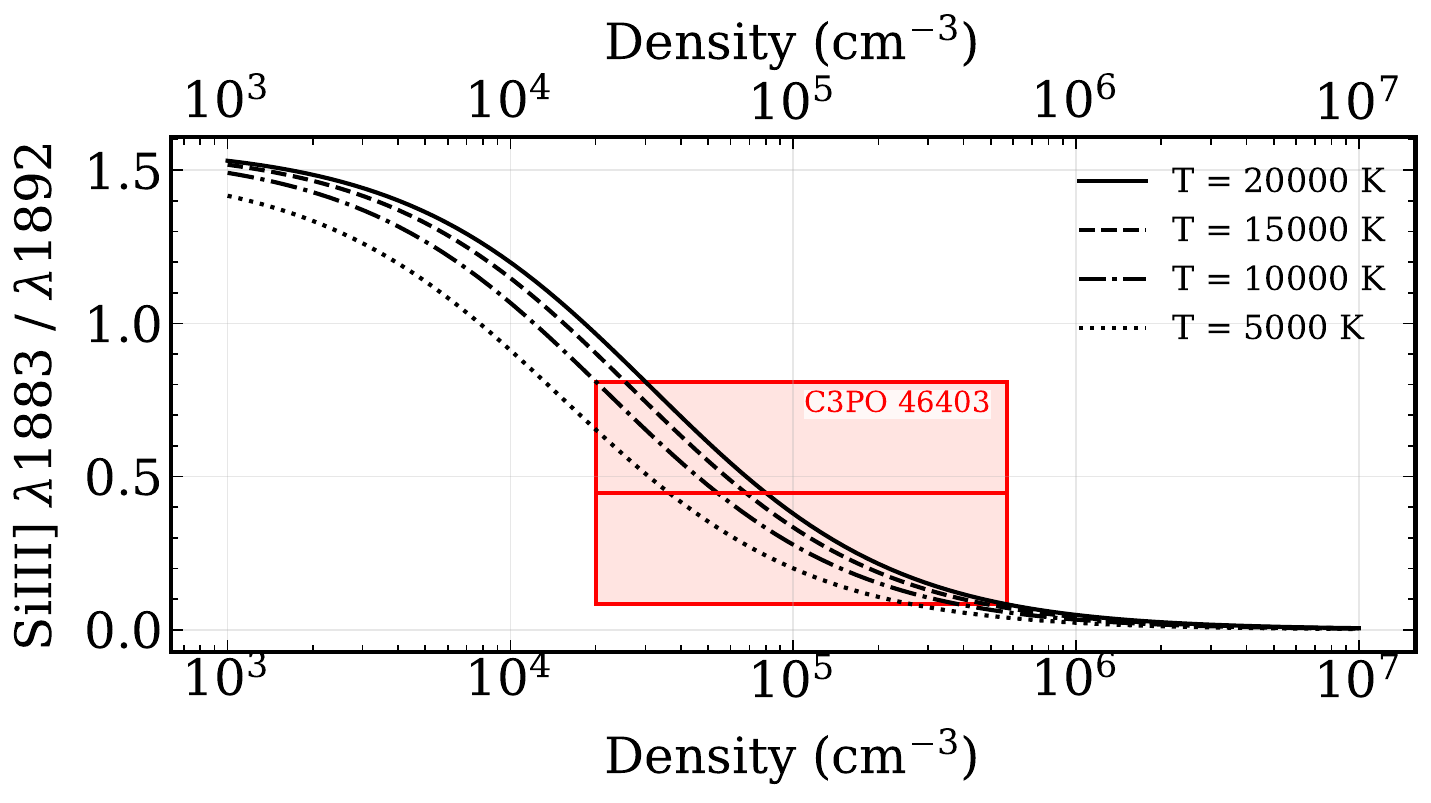}
    \includegraphics[width=0.95\linewidth, trim={0 48pt 0 0}, clip]{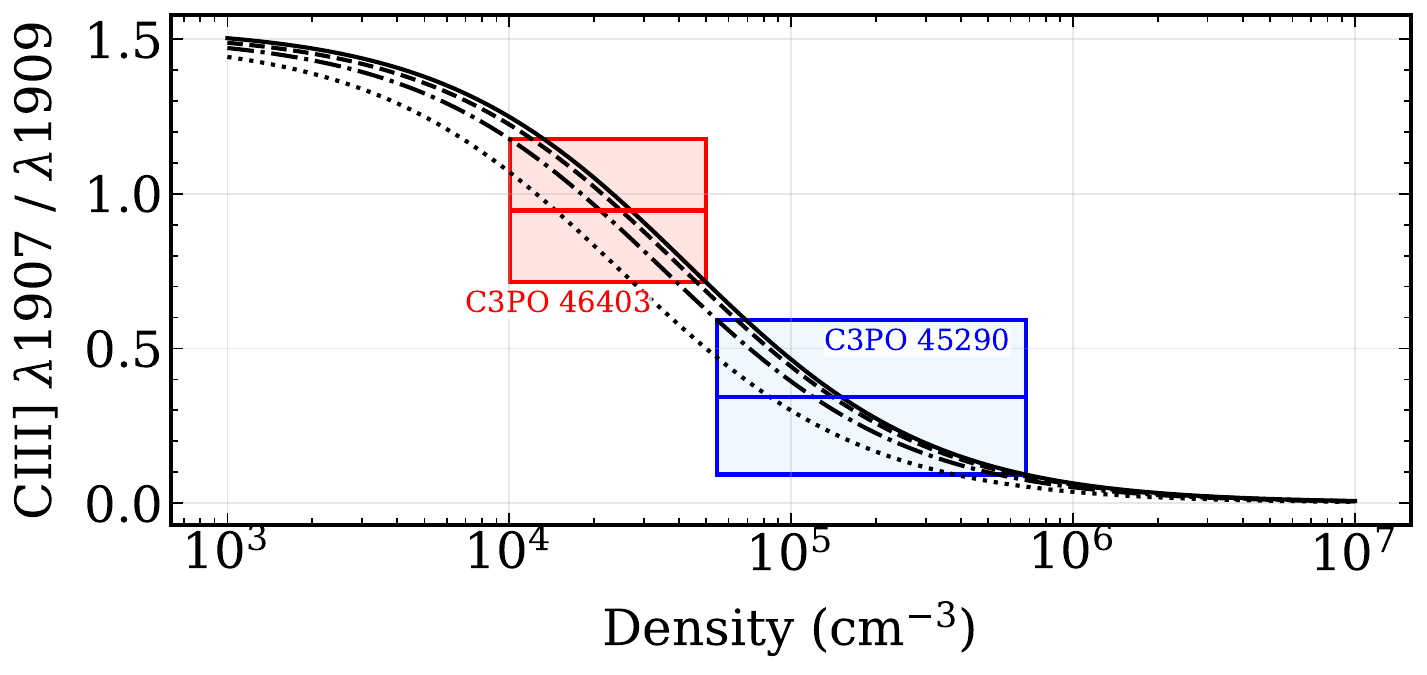}
    \includegraphics[width=0.95\linewidth]{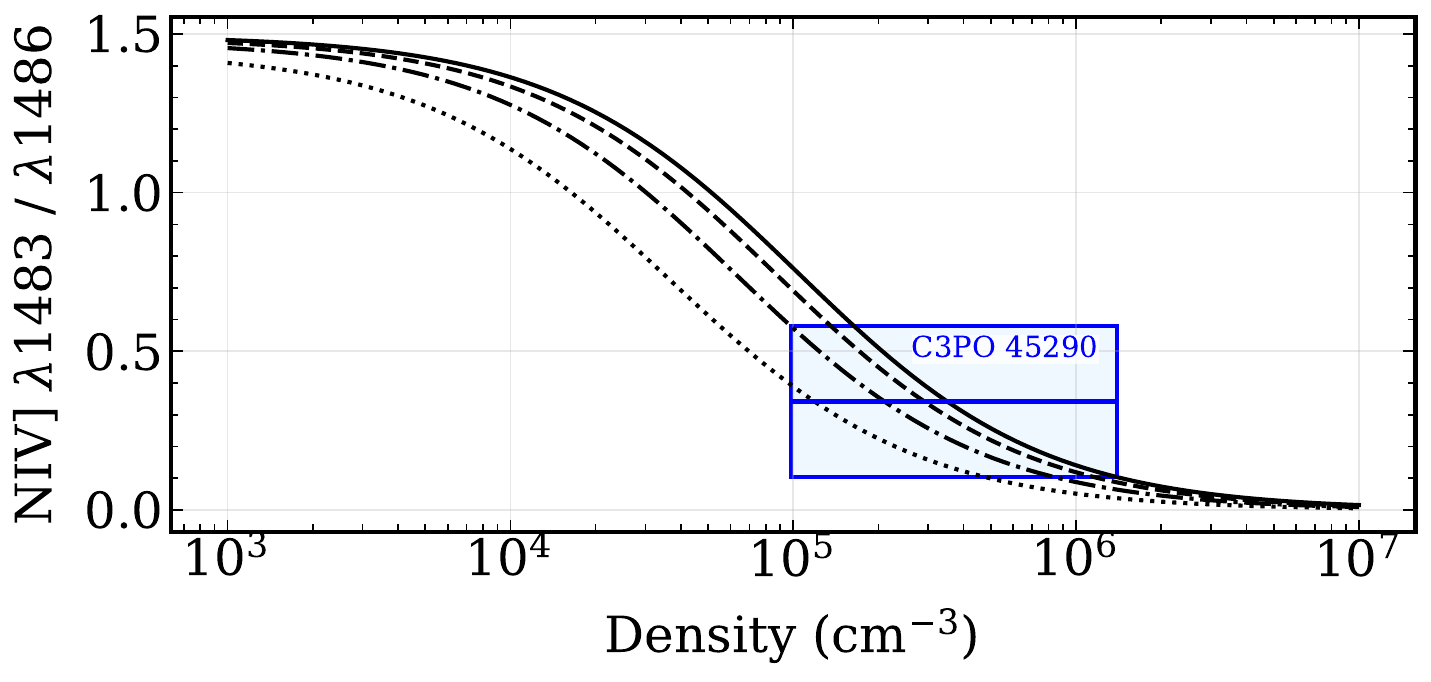}
    \caption{Relation between UV emission-line ratios and gas density.  The top panel shows  \ion{Si}{3}] $\lambda$1883 / \ion{Si}{3}] $\lambda$1892, the middle panel shows [\ion{C}{3}] $\lambda$1907 / \ciii\ $\lambda$1909, and the bottom panel shows \ion{N}{4}] $\lambda$1483 / \ion{N}{4}] $\lambda$1486.  The predicted curves correspond to different gas temperatures, $T$, as indicated in the legend.  In each panel, the blue and red boxes show the measure line ratios for C3PO 45290 and 46403, respectively (as labeled).  The height of the boxes corresponds to the $\pm 1\sigma$ uncertainties on the line ratio, and the width denotes the allowed range in density for gas temperatures of $10,000$ to $20,000$~K.}
    \label{fig:uvline_ratios}
\end{figure}

Figure~\ref{fig:uvline_ratios} shows the theoretical emission line ratios for \ion{Si}{3}] $\lambda$1883 / \ion{Si}{3}] $\lambda$1892, [\ion{C}{3}] $\lambda$1907 / \ciii\ $\lambda$1909, and \ion{N}{4}] $\lambda$1483 / \ion{N}{4} $\lambda$1486, calculated from \texttt{PyNeb} \citep{Luridian_2015} for gas temperatures of 5000 to 20,000~K.  The measured values for C3PO 43640 and 45290 are indicated by red and blue-shaded boxes, respectively.   In both galaxies, the implied gas densities associated with these narrow UV lines are high.  Each LRD has a high-significance measurement of the \ciii\ doublet, reinforced either by the \ion{Si}{3}] or \ion{N}{4}] doublets.   These lines probe ionization energies spanning $\sim$16--47~eV, where studies from local galaxies have shown that measurements of the electron density based on these line ratios are generally consistent \citep{Mingozzi_2022}. 

For C3PO 46403, the \ciii\ doublet has a ratio of  [\ion{C}{3}] $\lambda$1907 / \ciii\ $\lambda$1909 = 0.95 $\pm$  0.23.  Assuming a gas temperature of 20,000~K (see Section~\ref{section:abundances}), this yields a gas density of $\log n_e / \mathrm{cm}^{-3} = 4.3\pm 0.3$, with a shift of $\approx$0.1~dex for a gas temperature of 15,000 to 25,000~K.   For C3PO 46403 we also detect \ion{Si}{3}] $\lambda$1892, with a stringent limit on \ion{Si}{3} $\lambda$1883 (Figure~\ref{fig:46403_uvzoom}) such that the line ratio is \ion{Si}{3}] $\lambda$1883/\ion{Si}{3}] $\lambda$1892 = 0.45 $\pm$ 0.36.  Weaker emission from the \ion{Si}{3}] $\lambda$1883 line is expected for high density gas, and  Figure~\ref{fig:uvline_ratios} shows the ratio of \ion{Si}{3}] $\lambda$1883 / \ion{Si}{3}] $\lambda$1892 spans a density range of $\log n_e / \mathrm{cm}^{-3} \approx 4.3$ -- 5.6 for gas temperatures $T_e > 10,000$~K.  This is consistent with the gas density derived from the \ciii-based ratio. 

For C3PO 45290, we detect the \ciii\ line doublet, but as illustrated in Figure~\ref{fig:45290_uvzoom}, the [\ion{C}{3}] $\lambda$1907 line is significantly weaker than the \ciii\ $\lambda$1909 line. This yields a ratio of [\ion{C}{3}] $\lambda$1907 / \ciii\ $\lambda$1909 = 0.34 $\pm$  0.25.   For a temperature of 15,000~K (see Section~\ref{section:abundances}), the  ratio yields a high gas density, $\log n_e / \mathrm{cm}^{-3} = 5.1\pm 0.4$, again with a shift of $\approx$0.1~dex for a range of gas temperature from 10,000 to 20,000~K.   For C3PO 45290, we also detect the \ion{N}{4}] doublet (Figure~\ref{fig:45290_uvzoom}), with a  ratio \ion{N}{4}] $\lambda$1483 / \ion{N}{4}]$\lambda$ 1486 = 0.34 $\pm$ 0.24.  Figure~\ref{fig:uvline_ratios} shows this corresponds to a high gas density value, $\log n_e/\mathrm{cm}^{-3} = 5.0$ -- 6.1,  
consistent with the \ciii-derived value.

\begin{deluxetable}{cl|r|r}
\tablecolumns{4}
\tablewidth{0pt}
\tablecaption{Derived Gas Properties for C3PO LRDs\label{table:quantities}}
\tablehead{ 
& \multicolumn{1}{l}{Quantity} & \colhead{C3PO 45290} & \colhead{C3PO 46403}
}
\startdata
\multirow{6}{*}{\rotatebox{90}{narrow lines}} & 
 $T_e / \mathrm{K}$ & $< 16000$ & 20000 $\pm$ 4600 \\ 
 & $\log( n_e(\ciii) / \mathrm{cm^{-3}} )$ & 5.1 $\pm$ 0.4 & 4.3 $\pm$ 0.3 \\
 & 12 + log O/H & $>7.52$\tablenotemark{$\dag$} & $>7.55$\tablenotemark{$\ddag$} \\ 
 & log $U$ & $-1.5 \pm 0.3$ & $-1.7 \pm 0.1$ \\ 
 & log C/O\tablenotemark{$\ast$} & $-0.96 \pm 0.16$ & $-0.76 \pm 0.07$ \\
 & log N/O\tablenotemark{\S} & $-0.32 \pm 0.13$ & $0.03 \pm 0.18$ \\\hline
\multirow{4}{*}{\rotatebox{90}{broad lines}} & & \\ 
& \multirow{1}{*}{$T_e / \mathrm{K}$ (assumed)} & \multirow{1}{*}{15000--25000} & \multirow{1}{*}{15000--25000} \\
& \multirow{1}{*}{$\log( n_e(\oiii) / \mathrm{cm^{-3}})$} &  \multirow{1}{*}{7.4 $\pm$ 0.5} & \multirow{1}{*}{6.8 $\pm$ 0.5} \\
  & & \\
 \enddata
 \tablenotetext{$\dag$}{Limit assumes upper limit on electron temperature.}
\tablenotetext{$\ddag$}{Limit assumes O/H $>$ $\mathrm{O^{2+}/H}$, see Section~\ref{section:abundances}.}
\tablenotetext{*}{Derived from [\ion{C}{3}] $\lambda$1907 + \ciii\ $\lambda$1909 and \oiiiuv\ $\lambda\lambda$1661,1666 with an ICF for C$^{2+}$/O$^{2+}$, see Section~\ref{section:abundances}.}
\tablenotetext{\S}{Value for C3PO 45290 derived from \ion{N}{4}] $\lambda$1486 / \ion{O}{3}] with an ICF for N$^{3+}$/O$^{2+}$. Value for C3PO 46403 derived from \ion{N}{3}] $\lambda\lambda$1746,1748 / \ion{O}{3}] with an ICF for N$^{2+}$/O$^{2+}$, see Section~\ref{section:abundances}.}
\end{deluxetable}

Therefore, both LRDs, C3PO 46403 and 45290, show evidence of high gas densities in the ionized zone of the narrow line regions in these galaxies.  Table~\ref{table:quantities} summarizes these results. These densities fall in the range spanned by \ciii-based electron densities measured for other star-forming galaxies at low and high redshifts, see Section~\ref{section:discussion_structure}. 

\subsection{Broad-Line Components}\label{section:broad_gas}

\begin{figure}
    \centering
    \includegraphics[width=0.95\linewidth]{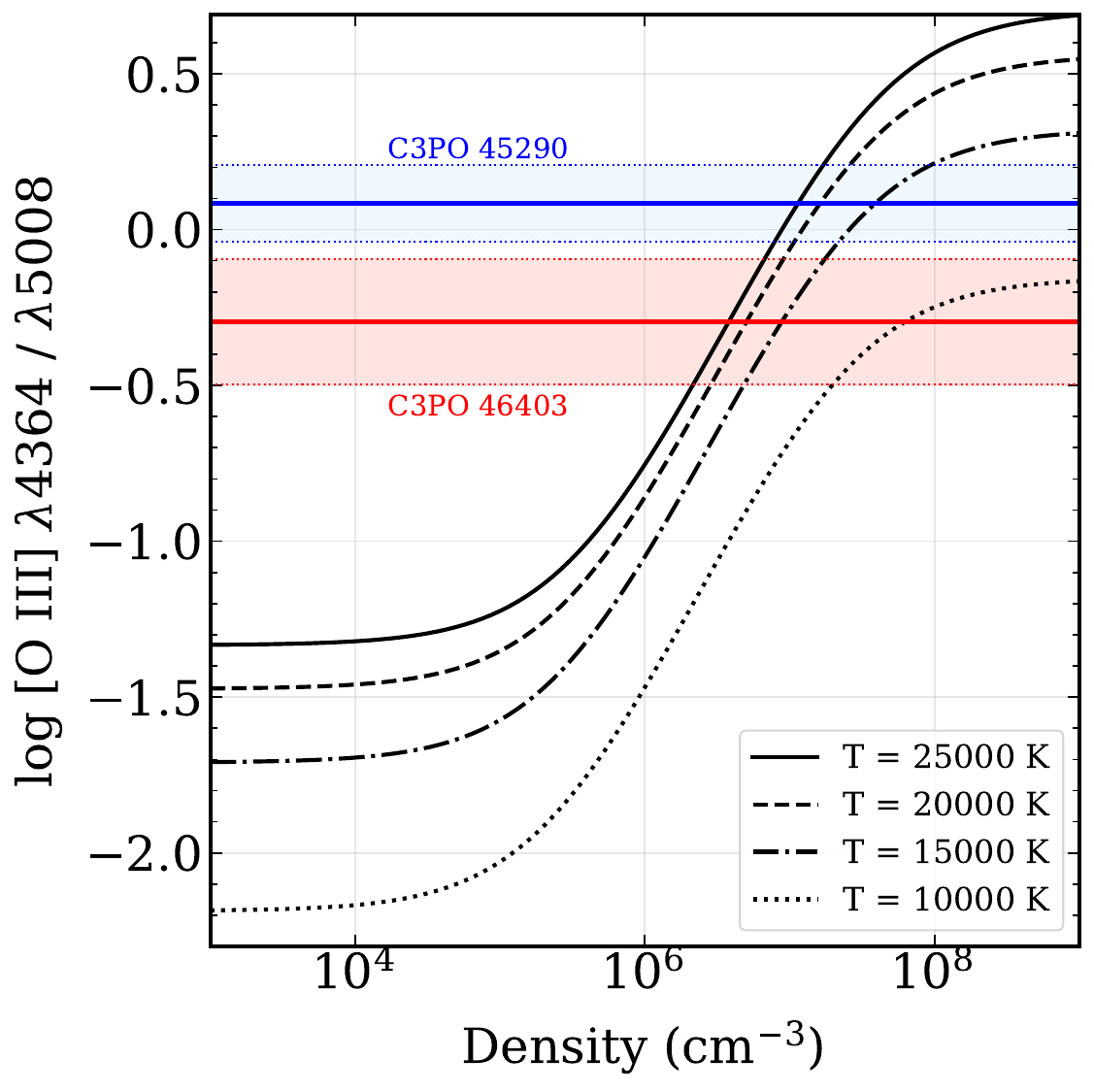}
    \caption{Relation between \oiii\ $\lambda$4364/\oiii\ $\lambda$5008  emission-line ratio and gas density.  The predicted curves correspond to different gas temperatures, $T$, as indicated in the legend.  The horizontal lines show the measured value of the line ratio of \oiii\ $\lambda$4364 / \oiii\ $\lambda$5008 for the broad components measured in C3PO 45290 and 46403 and their $\pm 1\sigma$ uncertainties, as labeled. }
    \label{fig:o3line_ratios}
\end{figure}

Both C3PO LRDs show evidence of broad \oiii\ $\lambda$4364 lines.  
%
%
Our emission-line fits yield measurements of the ratio of \oiii\ $\lambda$4364 / \oiii\ $\lambda$5008 in the narrow and broad components separately (Figures~\ref{fig:46403_balmerOiii} and \ref{fig:45290_balmerOiii}, and Table~\ref{table:optlines}).  The [\ion{O}{3}] $\lambda$4364 and \oiii\ $\lambda$5008 lines originate from two different excited metastable states with the same ground electron configuration, $1s^22s^22p^2$.  The \oiii\ $\lambda$4364 auroral line arises from the decay from a higher state to the intermediate state, $^1S_0 \rightarrow ^1D_2$.  The \oiii\ $\lambda$5008 line is produced when the intermediate state decays to the lower state, $^1D_2 \rightarrow ^3P_2$.   The critical densities of these lines differ because they originate from different upper energy levels with vastly different lifetimes. The \oiii\ $\lambda$5008 line decays via a magnetic dipole transition with a radiative probability that is relatively slow ($A \approx 0.02$~s$^{-1}$), while the \oiii\ $\lambda$4364 line decays via a forbidden electric quadrupole transition relatively quickly ($A \approx 1.6$~s$^{-1}$).  As the critical density is the density when the rate of collisional de-excitations equals the rate of spontaneous radiative decays, the critical density is much higher for the \oiii\ $\lambda$4364 line, $n_\mathrm{crit}(\oiii\ \lambda4364) \approx 3\times 10^7$~cm$^{-3}$, compared to the \oiii\ $\lambda$5008 line, $n_\mathrm{crit}(\oiii\ \lambda5008) \approx 7\times 10^{5}$ cm$^{-3}$.  At relatively low densities, $n \ll n_\mathrm{crit}(\oiii~5008)$, the ratio of the line emissivities is independent of density, and only depends on the gas electron temperature.  However, at higher gas densities, $n_\mathrm{crit}(\oiii~\lambda5008)$ $\lsim n \lsim$ $n_\mathrm{crit}(\oiii~\lambda 4364)$, the ratio becomes highly dependent on gas density.  

%

Considering only the broad components of the \oiii\ $\lambda$4364 and \oiii\ $\lambda$5008 lines, the ratios for the LRDs here are log~\oiii\ $\lambda$4364 / \oiii\ $\lambda$5008  = $0.084 \pm 0.122$ and  $-0.30\pm 0.20$ for C3PO 45290 and 46403, respectively. 
The fact that the flux of the broad \oiii\ $\lambda$4364 line is high compared to the flux in the broad component of \oiii\ $\lambda$5008 line for both LRDs is indicative of higher gas densities in these components.   Similar conditions have been observed in the broad-line regions of PG Quasars \citep{Boroson_1992}.  

Figure~\ref{fig:o3line_ratios} shows the expected ration of \oiii\ $\lambda$4364 / \oiii\ $\lambda$5008 as a function of temperature for a range of gas temperature, compared to these measured values.  Several previous studies of LRDs have noted high (total) \oiii\ $\lambda$4364 / \oiii\ $\lambda$5008 ratios \citep[e.g.,][]{Kokorev_2024b,Taylor_2025}.  Here, we have quantified this in terms of the broad-line components. While we have no direct measure of the gas temperature of the broad lines, they are expected to be $\sim$10,000--20,000~K \citep[see, e.g.,][]{Baskin_2005}.  Here, the broad-line \oiii\ $\lambda$4364 / \oiii\ $\lambda$5008 ratios correspond to gas densities of $\log (n_e/\mathrm{cm^{-3}}) = 6.8\pm 0.5$ to $7.4\pm 0.5$ for C3PO 46403 and 45290, respectively, assuming gas temperature $15,000-25,000$~K (Table~\ref{table:quantities}).  We stress these densities are likely lower limits.  For $T_e < 15,000$~K, the densities would be even higher.  Furthermore, \ion{He}{1} $\lambda$5017 could contribute to the flux we attribute to the broad \oiii\ $\lambda$5008 line, which may be important for C3PO 46403 where we observe strong \ion{He}{1} $\lambda$5877 (Figure~\ref{fig:46403_spec}). Similarly, in C3PO 46403 we attribute some of the \hgamma+\oiii\ $\lambda$4364 complex to [\ion{Fe}{2}] $\lambda\lambda$4351--8 (Figure~\ref{fig:46403_balmerOiii}).  Removing either the contribution of \ion{He}{1} or [\ion{Fe}{2}] from these lines would have the effect of \textit{increasing} the \oiii\ $\lambda$4364 / \oiii\ $\lambda$5008 ratio, \textit{increasing} the gas density.   Therefore, the \oiii--based gas densities of the broad lines are high, exceeding the densest measurements of broad-line regions of quasars (e.g., \citealt{Baskin_2005}).  We discuss this further below in Section~\ref{section:discussion_structure}. 

\section{Discussion}\label{section:discussion}

\subsection{Elemental Abundances and Rapid Enrichment in LRDs}\label{section:abundances}

The C3PO LRDs have evidence of low metallicity based on the oxygen emission-lines, but this is tentative because we have insufficient data for rest-optical oxygen lines to determine direct-temperature oxygen abundances.  C3PO 46403 lacks coverage of \oii, and has only a weak detection of the narrow-line component of \hbeta\ ($<5\sigma$).   C3PO 45290 has a very weak detection of the narrow-line component of the \oiii\ $\lambda$4364 line  ($<1\sigma$), which prevents obtaining a robust direct electron temperature.  Nevertheless, if we use the constraints on the \oiii\ lines from Table~\ref{table:optlines}, C3PO 45290 yields temperatures of $\log T_e / \mathrm{K}$ = $<4.2$ with 12 + log(O/H) $>$ 7.52 (less than half the Solar value).  For C3PO 46403, the narrow \oiii\ lines have robust detections, yielding a temperature of $\log T_e/\mathrm{K} = 4.3 \pm 0.1$.  However, we lack coverage of \oii, so we infer only O$^{2+}$/H = $(2.0\pm 0.5) \times 10^{-4}$.  This implies a total oxygen abundance of 12 + log(O/H) $>$ 7.55.  Therefore, the limits on the O/H abundances for both LRDs are consistent with being sub-Solar, but these remain inequalities. However, this conclusion is consistent with a recent analysis by \citet{Nikopoulos_2026} who find sub-Solar O/H abundances in other LRDs.

The UV emission lines provide evidence for low C/O in the C3PO LRDs. 
Using the \ciii\ and \oiiiuv\ lines we measure a C$^{2+}$/O$^{2+}$ ratio for each object using \texttt{PyNeb} assuming a temperature range $T_e = 15,000 - 20,000$~K.\footnote{While we use the default atomic data with \texttt{PyNeb} package to calculate ionic
abundances, we use the O$^{2+}$ collision strengths from \citet{Aggarwal_1999} as they include a six-level atom approximation, which is required to analyze the \oiiiuv\ $\lambda\lambda$1661,1666 lines.}   We take C/O $=$ C$^{2+}$/O$^{2+} \times \mathrm{ICF(C^{2+}/O^{2+})}$.  $\mathrm{ICF(C^{2+}/O^{2+})}$ is the ionization correction factor for these ions, which estimate from \citet{Berg_2019}. The ICF is dependent on the ionization parameter, $\log U$, which we estimate from the \oiii\ $\lambda$5008/\oii\ $\lambda$3727 ratio, and the metallicity \citep{Berg_2019}.  For C3PO 45290 we obtain, $\log U = -1.5 \pm 0.3$ using the values in Table~\ref{table:optlines}. For C3PO 46403, our G395M data do not cover \oii, so we take \oiii\ $\lambda$5008/\oii\ $\lambda$3727 from \citet{DEugenio_2026}, which yields $\log U = -1.7 \pm 0.1$.    The ICF$(\mathrm{C^{2+}/O^{2+})}$ values are then $1.30^{+0.26}_{-0.16}$ and $1.17^{+0.21}_{-0.20}$ for C3PO 45290 and 46403, respectively.   Applying these to the C$^{2+}$/O$^{2+}$ ratios, we obtain for C3PO 45290 $\log( \mathrm{C/O}) = -1.09 \pm 0.14$ assuming $T_e = 15,000$~K. For C3PO 46403 we obtain $\log \mathrm{C/O} = -0.82 \pm 0.10$ assuming $T_e = 20,000$~K.    These are listed in Table~\ref{table:quantities}.   

In both cases, the C/O abundances of the C3PO LRDs are at least $\lesssim$0.4~dex below Solar values reported in the literature, which span log (C/O)$_\odot\ = -0.44$ to $-0.14$ \citep{Anders_1989,Feltre_2016,Asplund_2021}.   These low C/O values are characteristic of other high-redshift objects \citep{Jones_2023,Hu_2024} with low metallicity. 

The detections of \ion{N}{4}] $\lambda$1486 in C3PO 45290 and \ion{N}{3}] $\lambda\lambda$1746,1748 in C3PO 46403 are intriguing.    
Detections of nitrogen UV emission lines are uncommon, and require enhanced nitrogen abundances or AGN ionization \citep{Hainline_2011,Alexandroff_2018,Sobral_2018}.  While the detection of nitrogen lines in itself could be evidence that ionizing radiation from the LRDs is escaping into the narrow-line regions, many low-metallicity, star-forming galaxies also show evidence for nitrogen UV emission lines, both at high redshifts \citep[e.g.,][]{Topping_2024,Isobe_2023,Cameron_2023,Cameron_2024,Marques-Chaves_2024,Maiolino_2024_nature,Morel_2026}, and low redshifts \citep[e.g.,][]{Mingozzi_2024}. As nitrogen production generally requires secondary channels \citep{Kobayashi_2020}, the nitrogen abundance is expected to be low for low-metallicity objects (see also \citealt{Cameron_2023,Senchyna_2024}).  The detection of the enhanced nitrogen in high redshift galaxies generally therefore implies rapid nitrogen-production channels \citep[e.g.,][]{Zhu_2026}.

We can use the detection of the \ion{N}{4}] lines in C3PO 45290 and \ion{N}{3}] in C3PO 46403 along with \oiiiuv\ $\lambda\lambda$1661,1666 to estimate N/O. 
%
%
For C3PO 46403, we take $\mathrm{N}/\mathrm{O} = \mathrm{N^{2+}/O^{2+}} \times \mathrm{ICF(N^{2+}/O^{2+})}$.  
%
%
We compute $\mathrm{N^{2+}/O^{2+}}$ from the observed \ion{N}{3}]/\ion{O}{3}] ratio from \texttt{PyNeb} with $T_e$=20,000~K and $\log n_e/\mathrm{cm}^{-3}$ = 4, which yields $\log \mathrm{N^{2+}/O^{2+}} = -0.09$.  We estimate ICF$(\mathrm{N^{2+}/O^{2+}})$ from \citet{Martinez_2025} with $\log U$ from above, which yields ICF$(\mathrm{N^{2+}/O^{2+}})$ = 1.17$^{+0.21}_{-0.20}$.    The N/O ratio is then $\log \mathrm{N/O} = +0.03 \pm 0.18$.  These quantities are listed in Table~\ref{table:quantities}.  Changing the electron temperature from 10,000 -- 25,000~K or gas density from $\log n_e/\mathrm{cm}^{-3} = 3 - 5$ changes N/O by $<0.1$~dex.  

For C3PO 45290, we must use $\mathrm{N}/\mathrm{O} = \mathrm{N^{3+}/O^{2+}} \times \mathrm{ICF(N^{3+}/O^{2+})}$, as we have only a detection of \ion{N}{4}].  We compute $\mathrm{N^{3+}/O^{2+}}$ from the observed \ion{N}{4}]/\ion{O}{3}] ratio from \texttt{PyNeb} with $T_e$=15,000~K, which yields $\log \mathrm{N^{3+}/O^{2+}} = -0.86$.  We again use \citet{Martinez_2025} to estimate $\mathrm{ICF(N^{3+}/O^{2+})}$, which yields log ICF($\mathrm{N^{3+}/O^{2+}}$) = $0.55^{+0.23}_{-0.18}$~dex.  This yields $\log \mathrm{N/O} = -0.32 \pm 0.13$, not including the uncertainty from the ICF.  These quantities are listed in Table~\ref{table:quantities}.   To shrink these uncertainties requires detections of additional nitrogen--emission lines to better constrain $\log U$ in the nebula.  

Regardless, both C3PO LRDs show evidence of enriched N/O compard to the Solar value, $\log (\mathrm{N/O})_\odot = -0.86$ \citep{Asplund_2021}.  Therefore, based on the \ion{N}{4}] and \ion{N}{3}] detections in the C3PO LRDs, they show evidence for N/O enrichment, by 0.5--0.9~dex, but these are dependent somewhat on the assumed ionization parameter, particularly for C3PO 45290.

Super-solar N/O ratios are rare at any redshift \citep[e.g.,][]{Marques-Chaves_2024}.  Low-redshift star-forming galaxies and Galactic \ion{H}{2} regions typically show nitrogen abundances within 0.5~dex of solar \citep[e.g.,][]{Izotov_2023}.  It is difficult to explain the simultaneous super-solar N/O and sub-solar C/O values, as seen in the LRDs here.  Chemical-evolution models can achieve this, but only for single bursts with a narrow span of ages, $\sim 3-4$~Myr which produce a high density of nitrogen-rich Wolf-Rayet (WN) stars compared to oxygen-rich (WO) or carbon-rich Wolf-Rayet (W5C) stars \citep[e.g.,][]{Charbonnel_2023,Marques-Chaves_2024,Tapia_2024,Senchyna_2024,Shi_2026}. \citet{Zhu_2026} argue that elevated N/O in high-redshift galaxies observed by \jwst\ requires enrichment from WN binaries, producing nitrogen-rich, oxygen- and carbon-depleted ejecta \citep[see also,][]{Senchyna_2024}.   It seems unlikely that such conditions would be common in the full population of LRDs, unless the LRD phase in galaxies is related to young, strong starbursts.  Alternatively, some studies have found a positive correlation between N/O and ISM density \citep{Arellano-Cordova_2025}, which may suggest that a combination of density and enrichment is a necessary requisite for LRDs.  Therefore, if high N/O are frequent among LRDs, then it may link both to the star-formation, chemical-enrichment histories and physical conditions needed for the onset of an LRD phase in galaxies.

\begin{figure*}
    \centering
    \includegraphics[width=\linewidth]{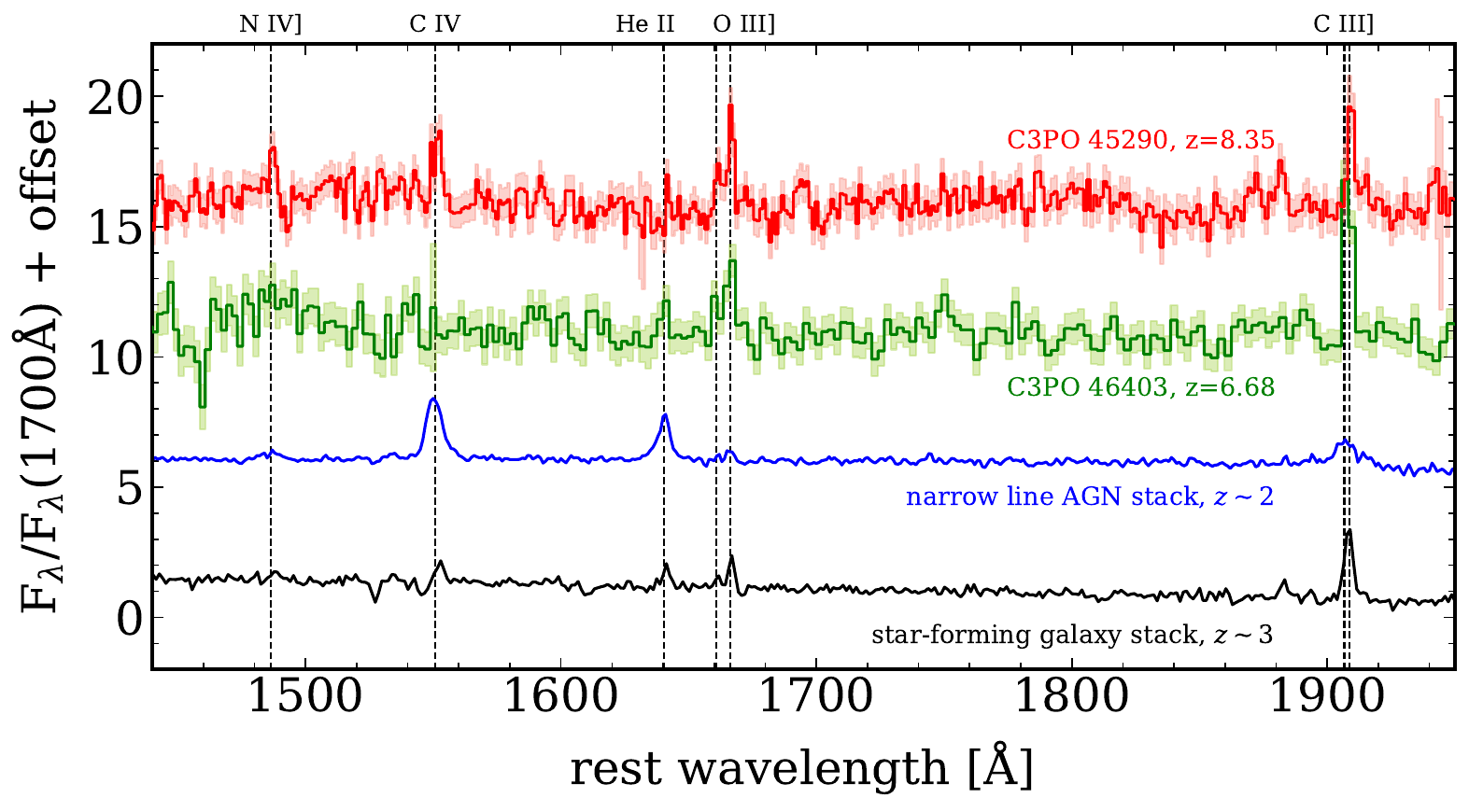}
    \caption{Comparison of rest-frame UV spectra of the C3PO LRDs and stacks of star-forming galaxies at $z\sim 3$ with $8 < \mathrm{EW(\ciii)} < 20$ \citep{Llerena_2022} and NLAGN at $z\sim 2$ \citep{Hainline_2011}.  The spectra cover the wavelength range that includes the UV diagnostic lines \ion{N}{4}], \civ, \heii, \oiiiuv, and \ciii.  Note that the spectral resolution of the C3PO LRDs is $R\sim 1000$, compared to $R\sim 700$ for the stacked NLAGN and star-forming galaxies.
    %
    %
    In contrast to the $z\sim 2$ NLAGN, the C3PO LRDs have weaker and narrower \heii, and stronger \oiiiuv\ and \ciii, more typical of star-forming galaxies, albeit with higher EW.}\label{fig:uv_spec_comp}
\end{figure*}

\begin{figure*}
    \centering
    \includegraphics[width=\linewidth]{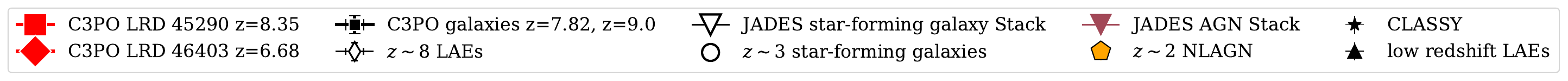}
    \begin{tikzpicture}
        \node[anchor=south west, inner sep=0] (image) at (0,0) {
            \includegraphics[height=2.5 in]{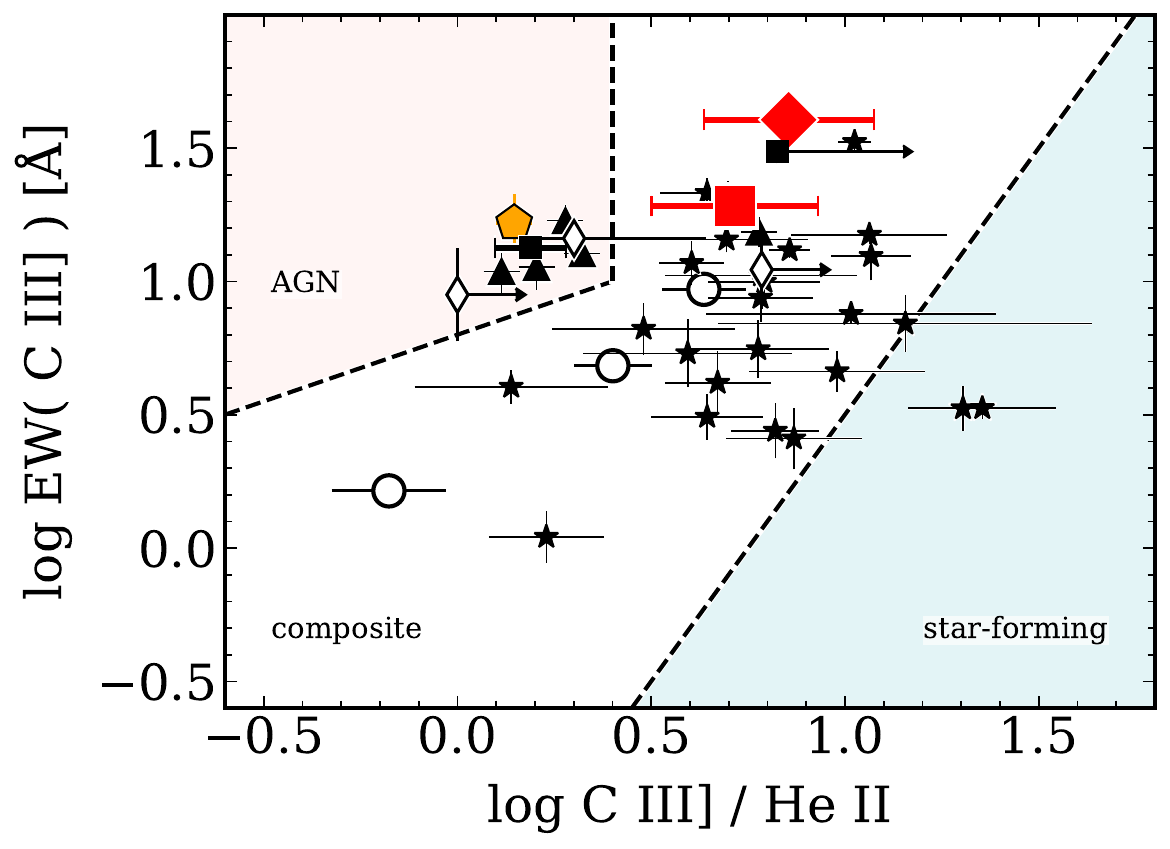}
        };
        \begin{scope}[x={(image.south east)},y={(image.north west)}]
            \node[anchor=north west, fill=white, fill opacity=0.1, text opacity=1,font=\large] at (0.2,0.98) {(A)};
        \end{scope}
    \end{tikzpicture}
    \hfill 
    \begin{tikzpicture}
        \node[anchor=south west, inner sep=0] (image) at (0,0) {
            \includegraphics[height=2.5 in]{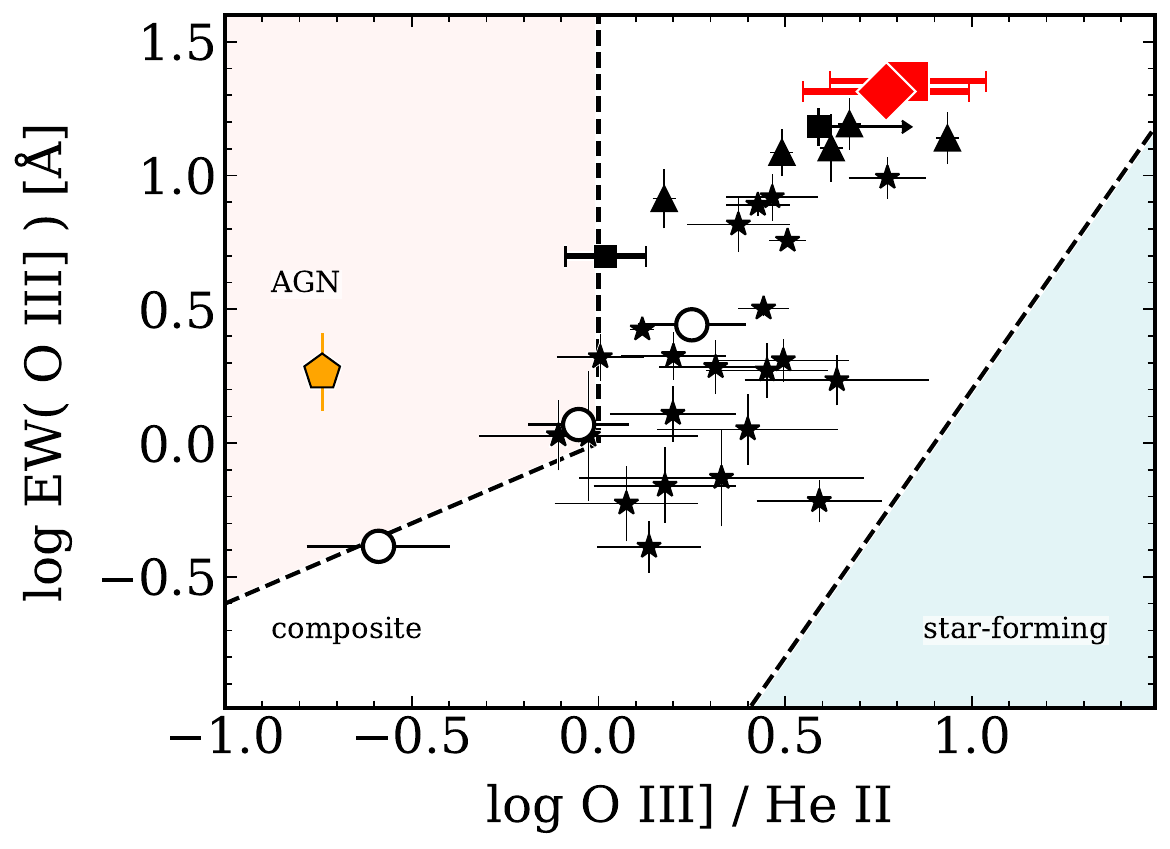}
        };
        \begin{scope}[x={(image.south east)},y={(image.north west)}]
            \node[anchor=north west, fill=white, fill opacity=0.1, text opacity=1,font=\large] at (0.2,0.98) {(B)};
        \end{scope}
    \end{tikzpicture}
    \begin{tikzpicture}
        \node[anchor=south west, inner sep=0] (image) at (0,0) {
            \includegraphics[height=2.5 in]{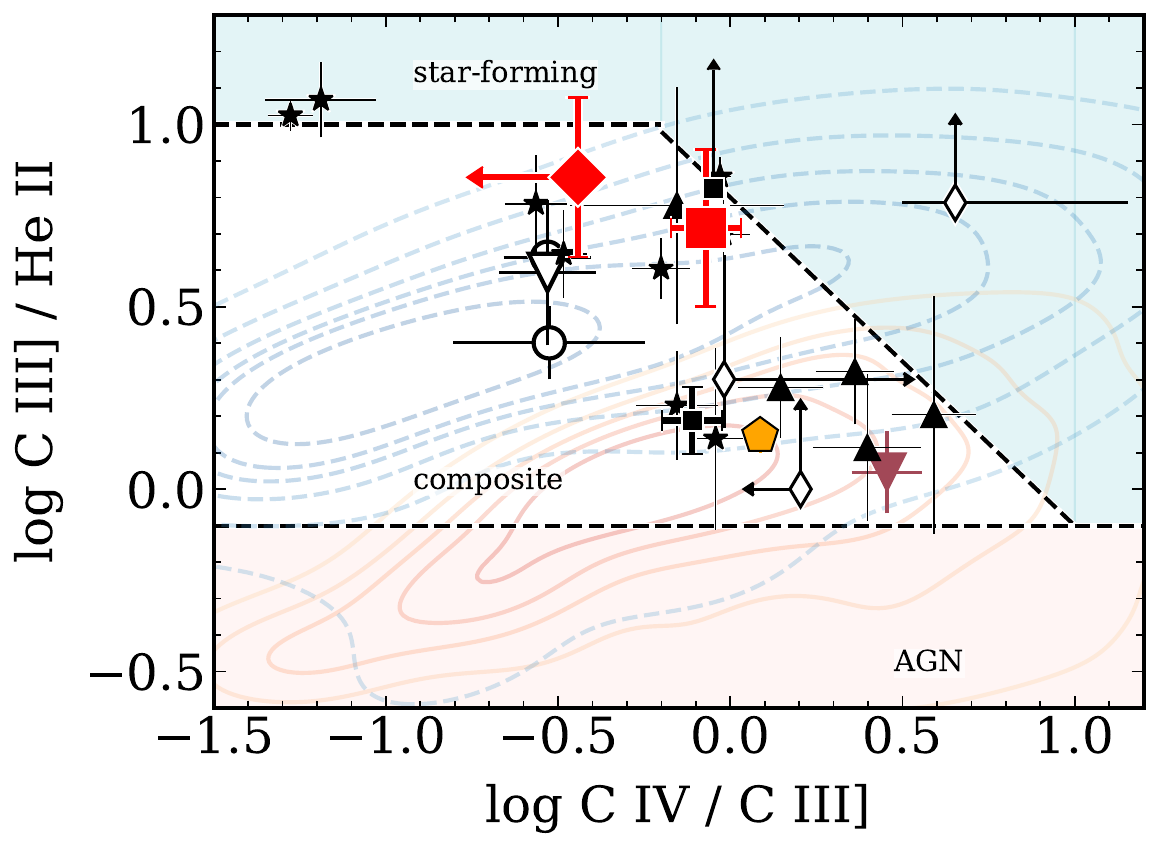}
        };
        \begin{scope}[x={(image.south east)},y={(image.north west)}]
            \node[anchor=north west, fill=white, fill opacity=0.1, text opacity=1, font=\large] at (0.2,0.98) {(C)};
        \end{scope}
    \end{tikzpicture}
    \hspace{14pt}
    \begin{tikzpicture}
        \node[anchor=south west, inner sep=0] (image) at (0,0) {
            \includegraphics[height=2.5 in]{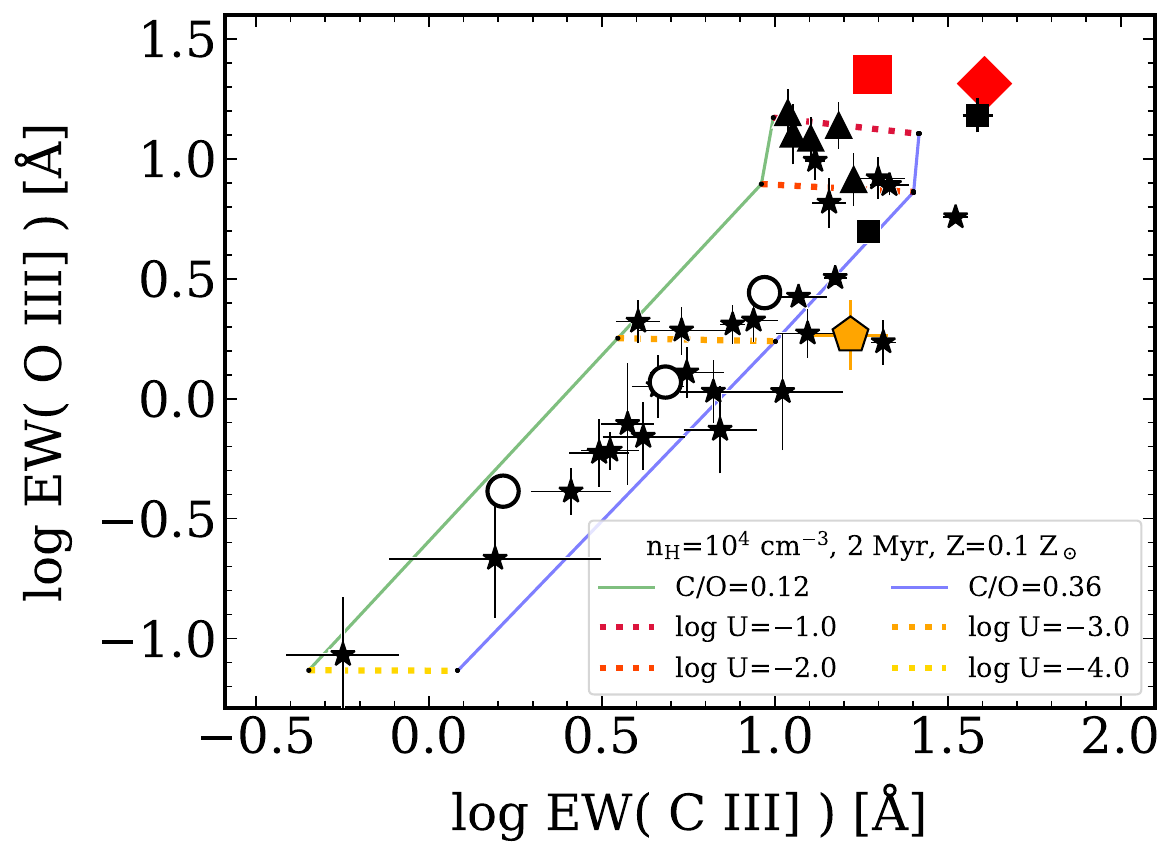}
        };
        \begin{scope}[x={(image.south east)},y={(image.north west)}]
            \node[anchor=north west, fill=white, fill opacity=0.1, text opacity=1, font=\large] at (0.2,0.98) {(D)};
        \end{scope}
    \end{tikzpicture}
    \hfill 

    \caption{UV emission line diagnostics. Panel (A) shows the \ciii/\heii\ ratio versus EW(\ciii)  for C3PO 46403 (red diamond) and C3PO 45290 (red square) compared to sources and predictions in the literature.  Panel (B) shows the \oiiiuv/\heii\ ratio versus EW(\oiiiuv).  Panel (C) shows the \civ/\ciii\ ratios versus \ciii/\heii\ ratios.  Panel (D) shows EW(\ciii) versus EW(\oiiiuv). The key to the data points is the same in each panel (see legend).  These include low--redshift star-forming galaxies from CLASSY (stars; \citealt{Mingozzi_2022}), low-redshift LAEs (upward triangles; \citealt{Jung_2025}), higher redshift galaxies at $z\sim 3$ from stacked spectra binned by EW(\ciii) (open circles; \citealt{Llerena_2022}) a stack of star-forming galaxies from JADES (open downward triangle; \citealt{Scholtz_2025}), and $z\sim 8$ LAEs (open diamonds; \citealt{Witstok_2025}).  The filled downward triangle and pentagon show stacks of AGN at $z\sim 2$ \citep{Hainline_2011} and from JADES \citep{Scholtz_2025}.  The black-filled squares show results from other C3PO galaxies at $z=8-9$ (Hu et al.\ in prep; Papovich et al.\ in prep). The shaded regions show areas designated as ``AGN'', ``star-forming'', and ``composite'' from \citet{Hirschmann_2019}.  In panel (C), the contours show predictions for star-forming galaxies with C/O $< 0.18$ \citep{Gutkin_2016} and AGN (assuming solar C/O, \citealt{Feltre_2016}).  In panel (D) the lines show a grid of models at fixed ionization parameter, $\log U$, and C/O ratio, for a stellar population with an age of 2 Myr, 0.1 $Z_\odot$, and gas density, $n_H=10^4$~cm$^{-3}$.  The C3PO LRDs show UV emission line ratios consistent with other star-forming galaxies at low and high redshifts, albeit the C3PO LRDs show stronger UV line EWs than most other galaxies. }
    \label{fig:uvline_diagnostics}
\end{figure*}

\subsection{On the Nature of the Emission-Lines in LRDs}

\subsubsection{Ionizing Source of the Rest-UV Narrow Lines}\label{section:ionzing_source}

We consider the possible origins of the rest-UV emission for the C3PO LRDs by comparing the spectra to other high-redshift objects. Figure~\ref{fig:uv_spec_comp} shows the rest-frame, 1500~\AA--1900~\AA, spectra for C3PO 46403 and 45290, the stacked (average) spectrum of star-forming galaxies with strong emission lines at $z\sim 3$ \citep{Llerena_2022}, and the stacked spectrum of narrow-line AGN (NLAGN) at $z\sim 2$ \citep{Hainline_2011}. The star-forming galaxy stack at $z\sim 3$ only includes galaxies with high EW(\ciii), 8--20~\AA, i.e., the most extreme emission-line galaxies in that sample. C3PO 45290 has EW(\ciii) $\simeq$ 20~\AA, at the upper bound of the values of galaxies in the $z\sim 3$ stack, and C3PO 46403 has EW(\ciii) $\simeq$ 40~\AA, significantly higher.  The latter is consistent with the measurement from \citet{Arevalo-Gonzalez_2026} for this galaxy based on NIRSpec prism data.   C3PO 45290 has strong \civ, comparable to the $z\sim 2$ NLAGN, but at much lower EW which is more consistent with the galaxies in the $z\sim 3$ stack. In contrast to both of the stacked spectra, the C3PO LRDs have higher EW(\ciii), higher EW(\oiiiuv), and high \oiiiuv/\heii\ ratios. These results suggest the rest-UV spectra of the C3PO LRDs area qualitatively more similar to the stacked $z\sim 3$ extreme emission-line star-forming galaxies, but the strength of the \ciii\ and \oiiiuv\ lines suggest that an additional source of ionization is required. 

The high \ciii/\heii\ and \oiiiuv/\heii\ ratios in both C3PO LRDs are indicative of star-formation being a primary source of photoionization of the UV narrow-lines with some contribution from a central engine of the LRD. 
%
%
Figures~\ref{fig:uvline_diagnostics}a and \ref{fig:uvline_diagnostics}b show the relation between the \ciii/\heii\ and \oiiiuv/\heii\ ratios and their respective equivalent widths.  Both C3PO LRDs fall in a region designated as ``composite'' as defined by \citet{Hirschmann_2019}, where both ionization from AGN and star-formation are able to produce the line ratios and EWs. This is not a conclusive determination given that nearly all of the extreme star-forming galaxies from the literature reside in the ``composite'' region, including those in the local universe \citep{Mingozzi_2022,Jung_2025}, at moderate redshifts, $z\sim 3$ \citep{Llerena_2022}, and at higher redshifts, observed by \jwst\ \citep[Hu et al.\ in prep., Papovich et al.\ in prep.]{Witstok_2025}.
In contrast, the stacked composite spectra of NLAGN at $z\sim 2$ has line ratios and EW values that place it fully in the AGN region of the \ciii/\heii\ and \oiiiuv/\heii\ EW diagrams, indicating that such objects have different ionizing fields compared to extreme star-forming objects including the C3PO LRDs.  Interestingly, the two C3PO LRDs occupy a distinct part of the ``composite'' region in \ref{fig:uvline_diagnostics}b with high EWs. 

The interpretation of Figures~\ref{fig:uvline_diagnostics} requires consideration of the abundance ratios of the high-redshift galaxies.   The regions designated ``AGN'', ``composite'', and ``star-forming'' are based on stellar-population models that span a range of metallicity from super-- to sub--solar \citep{Feltre_2016,Gutkin_2016,Hirschmann_2019}, with either a (fixed) Solar C/O ratio, \citep[C/O = 0.44][]{Feltre_2016} or a range of C/O \citep{Hirschmann_2019}.  However, observations of extreme emission line galaxies show they favor lower oxygen abundances of 5--20\% Solar, and sub-solar C/O ratios of $0.1-0.2$ \citep[less than half Solar, e.g.,][J.\ Yang et al. in prep., and references therein]{Jones_2023,Hu_2024}.  There is evidence this extends to the host galaxies of LRDs as well \citep{Nikopoulos_2026}.  

Our C3PO measurements favor reduced C/O values in both LRDs (Section~\ref{section:abundances}).  This has an impact as it lowers both the predicted \ciii/\heii\ ratios and EW(\ciii) values of nebular regions \citep[J. Yang et al.\ in prep.]{Jaskot_2016}. 
%
%
%
In Figure~\ref{fig:uvline_diagnostics}c the blue contours show the star-forming models of \citep{Gutkin_2016} restricted to C/O $<$ 0.18, equivalent to 50\% the Solar value (assuming (C/O)$_\odot$ = 0.36 \citealt{Anders_1989}), which is consistent with the C/O measurements of our LRDs (Section~\ref{section:abundances}).  In this case, the model predictions for star-forming galaxies shift almost entirely into the ``composite'' region.  These encompass both the literature star-forming sources and the C3PO LRDs. In contrast, the line ratios of galaxies selected to be high--redshift AGN \citep{Hainline_2011,Witstok_2025} fall nearer the ``AGN''/``composite'' boundary and overlap better with the predictions for AGN \citep[red contours,][]{Feltre_2016}.  It is noteworthy that these AGN models have C/O = (C/O)$_\odot$, and lowering this would lower the \ciii/\heii\ ratios while leaving the \civ/\ciii\ relatively unchanged (J.\ Yang et al.\ in prep.). 

While we have not corrected for dust attenuation, this has a minimal impact on our results. Even for $A(V)=0.7$ (see Section~\ref{section:analysis}), the corrected  \civ/\ciii\ ratio shifts by $\approx +0.07$~dex (higher), while the corrected \ciii/\heii\ and \oiiiuv/\heii\ ratios shift by $\approx -0.05$ and $-0.01$ (lower), assuming the \citet{Calzetti_2001} attenuation law. 

Regardless, we conclude that the \ciii/\heii\ and \civ/\ciii\ emission line ratios of the C3PO LRDs do not necessarily require ionization from AGN, and are consistent with predictions for star-forming galaxies with lower C/O. 

%

The high EW(\ciii) and EW(\oiiiuv) are peculiar attribute of the C3PO LRDs as illustrated in Figure~\ref{fig:uvline_diagnostics}d.  Indeed, the EW(\oiiiuv) of the LRDs are the highest of the comparison samples, with values $\geq 15$~\AA.  Similarly, C3PO 46403 has EW(\ciii) = 40~\AA, the highest among all the comparison samples. These values exceed expectations from star-forming models \citep{Jaskot_2016}.  The grid in Figure~\ref{fig:uvline_diagnostics}d shows predictions from \texttt{Cloudy} using BPASS stellar-population models with high gas density, $n_\mathrm{H}=10^4$~cm$^{-3}$, and a fixed metallicity, $Z=0.1$~$Z_\odot$ (J.\ Yang et al., in prep) and a range of ionization parameter, $\log U$, and C/O.  The LRDs lie at the extreme end, beyond even extreme Lyman-$\alpha$ emitters in the local Universe \citep{Jung_2025}. Lowering the gas densities or changing the metallicity in the \texttt{Cloudy} models has the effect of \textit{lowering} the EW(\ciii) and EW(\oiiiuv) (see also, e.g., \citealt{Jaskot_2016}). Therefore, to produce the EW of the UV emission lines an additional source of high-energy photons is required that is presumably associated with the LRD accretion disk.  

\begin{figure}
    \centering
    \includegraphics[width=1\linewidth]{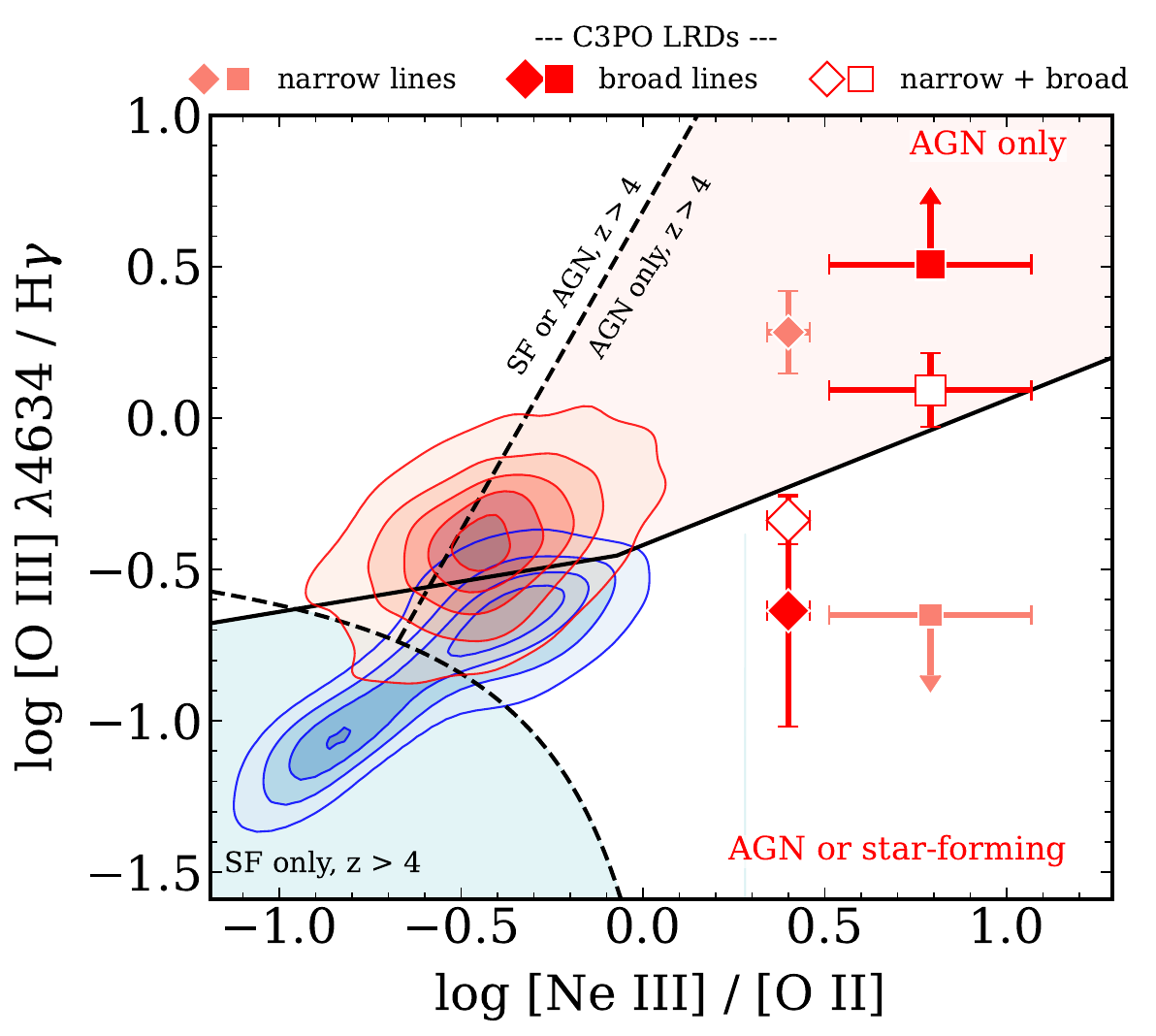}
    \caption{[\ion{Ne}{3}]/\oii\ versus \oiii\ 4364/\hgamma\ diagram as a diagnostic of AGN ionization.  The solid line indicates the region separating sources with ionization from AGN only from the region occupied by AGN or star-forming galaxies from \citet[red labels indicate regions of ``AGN only'' or ``AGN or star-forming'']{Mazzolari_2024}. The dashed lines show regions at $z > 4$ occupied by star-forming (SF) galaxies only, SF or AGN, or AGN only, favored by \citet[black labels]{Backhaus_2025}.  The red and blue contours show galaxies at $z\sim 0.1$ selected as AGN and star-formation, respectively, based on their optical emission lines from SDSS DR7 (see text).  The squares and diamonds show measurements of LRDs, C3PO 45290 and 46403, respectively, based on our measurements of the narrow lines, broad lines, and the sum of the two components (as labeled). Both C3PO LRDs show evidence that AGN contribute or dominate the ionization of these lines.}
    \label{fig:o3hg}
\end{figure}

\subsubsection{Ionizing Source of the Rest-Optical Lines}\label{section:ionizing_source_optical}

Rest-optical emission-line ratios of LRDs largely show they are typical of other star-forming galaxies \citep[e.g.,][]{Kocevski_2023,Harikane_2023}.  One exception is where LRDs fall in the diagram of [\ion{Ne}{3}]/\oii\ versus \oiii\ $\lambda$ 4364/\hgamma, which has been introduced as a means to separate AGN from star-forming populations \citep{Mazzolari_2024,Backhaus_2025}.  Several recent studies have employed the [\ion{Ne}{3}]/\oii\ versus \oiii\ $\lambda$ 4364/\hgamma\ diagram to diagnose the ionizing conditions in LRDs \citep{DEugenio_2026,Sok_2026}, which show LRDs favor a contribution of AGN to the ionization of these emission lines. 

Figure~\ref{fig:o3hg} shows [\ion{Ne}{3}]/\oii\ versus \oiii\ $\lambda$4364/\hgamma\ for the C3PO LRDs. Here, we take [\ion{Ne}{3}] as the flux of the [\ion{Ne}{3}] $\lambda$3870 line, and \oii\ as the sum of the flux of the \oii\ $\lambda\lambda$3727,3730 doublet.  The contours in Figure~\ref{fig:o3hg} show galaxies selected from the SDSS DR8 MPA-JHU value-added catalog \citep{Brinchman_2004} as AGN or star-forming using the the [\ion{N}{2}]/\ha\ versus \oiii/\hbeta\ emission line ratios \citep{Kewley_2001}.   For the C3PO LRDs, we show the line ratios separately for the narrow, broad, and total flux for the \oiii\ $\lambda$4364/\hgamma\ lines in the figure.  We show only the total [\ion{Ne}{3}]/\oii\ as neither of these lines show evidence for a broad component. For C3PO 46403, our G395M data do not cover \oii, so we adopt log [\ion{Ne}{3}]/\oii\ = 0.40 $\pm$ 0.06 from \citet{DEugenio_2026}.  

Both C3PO LRDs show evidence that AGN contribute to or dominate the ionization of these lines.  This is more evident in the narrow and total measurements for C3PO 45290, and in the narrow line emission for C3PO 46403.  In C3PO 45290, the broad \hgamma\ component is very weak, which yields log \oiii\ $\lambda$4364/\hgamma$|_\mathrm{broad} > 0.5$.  Based on modeling predictions \citep{Mazzolari_2024,Backhaus_2025} this requires ionization from an AGN, and indicates some direct illumination of the broad-line region from the LRD AGN accretion disk in this object.   This conclusion does not change if we consider the total line ratio for C3PO 45290, \oiii\ $\lambda$4364/\hgamma$|_\mathrm{total} > 0.1$.   It is only the narrow components for C3PO 45290, where the \oiii\ $\lambda$4364/\hgamma\ lines are consistent with model predictions for either star-formation or AGN.  

In C3PO 46403, the ratio of the narrow lines is log \oiii\ $\lambda$4364/\hgamma$|_\mathrm{narrow}$ = 0.28 $\pm$ 0.13, which requires ionization from an AGN.  While the ratio of the total lines and the ratio of the broad components place it in the region consistent with model predictions for star-formation or AGN, we note that the fluxes of the broad and narrow emission line components are very (\textit{anti-})correlated (Figure~\ref{fig:46403_balmerOiii}).  These results agree with measurements from \citet{DEugenio_2026} based on independent data (but they only consider a narrow-line origin for \oiii\ $\lambda$4364).  Regardless, this indicates that at least \textit{some} of the ionizing radiation incident on the \oiii\ $\lambda$4364 region stems from an AGN.

Therefore, for the C3PO LRDs we conclude that some ionizing radiation from an accretion disk appears necessary to contribute to the ionization of the \oiii\ $\lambda$4364 lines and the narrow UV lines. This requires that the dense gaseous medium surrounding the central engine of the C3PO LRDs has a non-unity covering factor such that this radiation can escape. This conclusion is similar to  other recent studies focusing on the \heii\ $\lambda$4687/\hbeta\ and \ion{O}{1}/\ha\ line ratios in LRDs \citep{Sok_2026,Wang_2026}, and from the velocity profiles of Lyman-$\alpha$ emission observed in some LRDs \citep{Tang_2026}. We return to the non-unity covering factor in the discussion below.

\begin{figure}
    \centering
    \includegraphics[width=\linewidth]{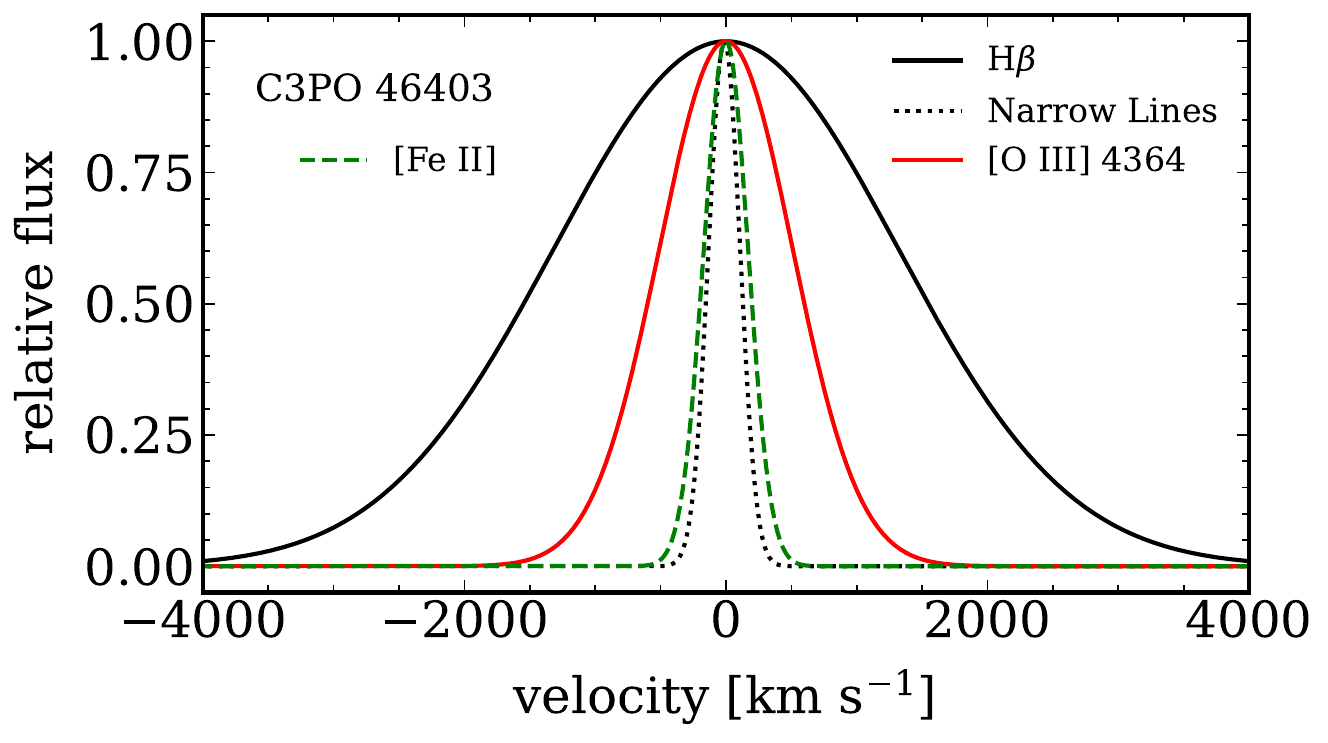}
    \includegraphics[width=\linewidth]{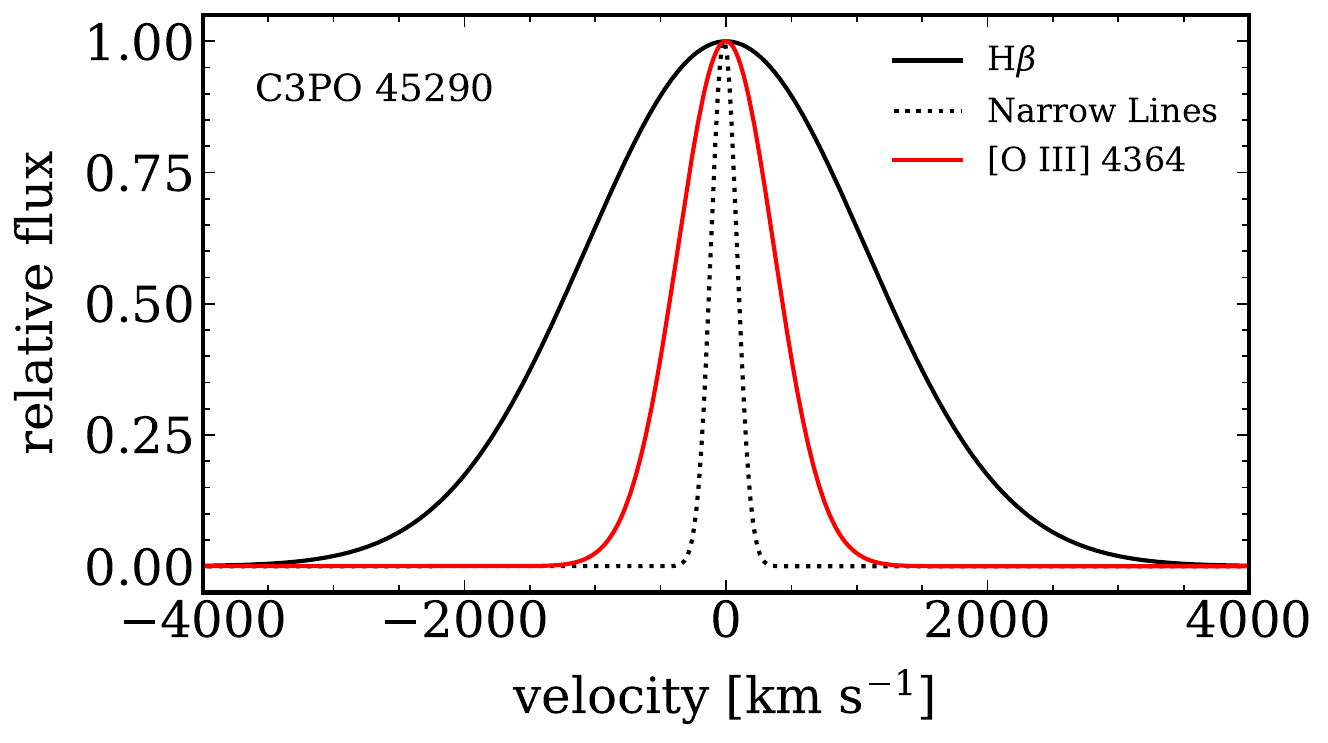}
    \caption{Comparison of emission line profiles for the broad \hbeta, \oiii 4364, and other lines, as labeled}
    \label{fig:line_profiles}
\end{figure}

\subsection{Structure of the Gaseous Regions of LRDs}\label{section:discussion_structure}

\subsubsection{Gas Kinematics and Spatial Sizes}

The emission line velocity widths provide information about the gas kinematics and geometry structure of the gas in LRDs.  Figure~\ref{fig:line_profiles} shows the fits to the emission-lines profiles in C3PO 46403 and 45290.  The Balmer lines, here illustrated by \hbeta, have FWHM $\approx$ 3000 km s$^{-1}$ for both objects.  The broad \oiii\ $\lambda$4364 lines in contrast are less broad with FWHM $\approx$ 1000 km s$^{-1}$, but are broader than the narrow lines. The origin of the broad lines observed in our LRDs can be attributed to either virial motions or outflows, both of which we consider below. 

If the velocities of the broad lines trace virial motions, we can estimate the sizes of the regions responsible for the different gaseous components.  We take as the size of the region that produces the narrow lines to be the effective radius, $R_e$, measured for the galaxies in the F150W band \citep{McGrath_2026}.  We use F150W because it probes the rest-UV in both galaxies where the emission from the central engine of the LRD has the least impact (see, Section ~\ref{section:ionzing_source}), while probing wavelengths redward of the Lyman-$\alpha$ break.  We can then estimate the sizes using FWHM$^2$ $\sim \sigma^2 \sim M R^{-1}$, yielding, 
\begin{equation}\label{equation:radius}
    R \lesssim R_e\left( \frac{\mathrm{FWHM}_n}{\mathrm{FWHM}} \right)^2,
\end{equation}
where FWHM corresponds to a given line width and where FWHM$_n$ is the width of the narrow lines, which we assume corresponds to the spatial dimension measured by $R_e$.  We use the inequality in Equation~\ref{equation:radius} because the FWHM$_n$ of the narrow component has not been corrected for the line-spread function due to the widths being roughly at the resolution. Therefore, our measured values of $R$ are considered an upper limit.  

Using the measurements from \cite{McGrath_2026}, we calculate circularized effective radii, $R_e = a_e\sqrt{q}$, where $a_e$ and $q=b/a$ are the effective semi-major axis and axis-ratio, respectively.   For C3PO 45290, the circularized effective radius is $R_e = 19 \pm 15$~pc.  For C3PO 46403, the circularized effective radius is $R_e=190\pm 84$~pc\footnote{\cite{DEugenio_2026} report a circularized effective radius of 126~pc based on independent fitting in the F200W band, consistent with our value.  We favor using the larger effective radius from F150W to provide a more secure size of the narrow-line region.}.  
This provides limits on the region of the broad, Balmer--emission region of 1.6$\pm$0.7~pc and $0.16 \pm 0.13$~pc for C3PO 46403 and 45290, respectively.\footnote{\citet{Lambrides_2026} estimated a size of the narrow-line regions in C3PO 46403 of $\sim 3\times 10^{16}$~cm $\approx 0.1$~pc, based on variability of the narrow \oiii\ $\lambda\lambda$4960,5008 lines. Using this value for the size of the narrow-line regions would reduce the region of the broad-lines by a factor of 1000.}  The region of the broad-\oiii\ $\lambda$4364 is larger, by approximately an order of magnitude, $12\pm 5$~pc and $1.4\pm 1.1$~pc, respectively.  Therefore, if the velocity width of the broad \oiii\ $\lambda$4364 lines as virial motions, then they probe scales within $\simeq$1--10~pc of the LRD center. 

One caveat to this discussion is that the broad Balmer emission-line components in LRDs may be broadened by electron scattering in the warm, partially ionized layers of the gaseous cocoon \citep[see, e.g.,][]{Chang_2026, Rusakov_2026,Torralba_2026}.   In this case, the density constraints yield size limits on the envelope of the LRD \citep{Rusakov_2026}, assuming $R \sim 3 N/n_e$, where $N$ is the column density of gas.   If the LRD accretion disks have strong intrinsic X-ray emission, then the  lack of X-ray emission requires $N \simeq 10^{24}$~cm$^{-2}$, consistent with analyses of the wings of the broad Balmer lines \citep{deGraaff_2026,Rusakov_2026,Sneppen_2026}.  Assuming the densities of the broad \oiii\ $\lambda$4364 lines are a lower limit for the conditions at smaller radii yields an upper limit on the LRD envelope of $R \lesssim 20000$~AU (about 0.1 pc), consistent with the values above.  Another reason for the smaller sizes here compared to those using the virial relation above is that the latter have been uncorrected for the instrumental LSF.  Even if the intrinsic width of the narrow lines is $\sim$3$\times$ smaller than measured, the virial sizes of the broad \oiii\ $\lambda$4364 region would be consistent with the implied densities for the LRD gaseous envelope, $n_e > 10^8$~cm$^{-3}$, derived from models of the LRD SED \citep{Naidu_2025,Taylor_2025,Ronayne_2026,Torralba_2026} and values derived from the wings of broad Balmer lines \citep{Kokorev_2026,Rusakov_2026,Sneppen_2026}.  

%
%

One intriguing note is that the [\ion{Fe}{2}] line widths of C3PO 46403 are narrower, with FWHM $\simeq$400 km~s$^{-1}$, implying that if they trace virial motions, then they originate from dense clouds further from the LRD center than the broad \oiii\ $\lambda$4364 component. This conclusion may be challenging to reconcile with models where the [\ion{Fe}{2}] lines originate in intermediate-density zones within $\sim$1~pc of the LRD center \citep[assuming BH* conditions,][]{Torralba_2026}, which would place the [\ion{Fe}{2}] emitting gas nearer the central engine than the broad \oiii\ components with larger line widths. 

Alternatively, the broad-line widths seen in the Balmer and/or \oiii\ $\lambda$4364 lines could indicate complex gas outflows.  Outflows of warm, partially ionized gas are seen as broad components in lines like \oiii\ $\lambda$5008, \ha, and [\ion{S}{2}] both at low and moderate redshifts \citep[see,][]{Wylezalek_2020,Weldon_2024}.  In these outflows, the broad emission components are typically blue-shifted from the galaxy systemic velocities by a few hundred km s$^{-1}$.   The gas densities of these stellar-driven outflows are typically $n_e \simeq 1000$~cm$^{-3}$ \cite[e.g.,][]{Weldon_2026} and those of AGN-driven outflows are typically $\log n_e/\mathrm{cm^{-3}} = 3.4-4.8$ \citep{Rose_2018},  
much lower than the gas densities we infer from \oiii\ $\lambda$4364 / \oiii\ $\lambda$5008.   Therefore, if the broad \oiii\ $\lambda$4364 lines represent outflows, these are much denser than that inferred from star-forming regions or AGN.  However, if the broad \oiii\ is due to an outflow, this could be an indication that the dense gas envelope of the LRD is being expelled, perhaps leading these sources to become more ``typical'' AGN.  This hypothesis merits further testing with future studies. 


Interestingly, there is no evidence for an offset between the centroid of the broad Balmer lines, particularly \hbeta, compared to the systemic velocity derived from the narrow lines (see Table~\ref{table:optlines}), which could be expected if the broad lines indicate outflows.  For C3PO 46403, the broad \oiii\ $\lambda$4364 is too blended with the other lines for us to detect any evidence of an offset of its centroid from systemic.  For C3PO 45290, the broad \oiii\ $\lambda$4364 is slightly offset from systemic by $(-106 \pm 67)$ km s$^{-1}$, but the significance is weak, $\simeq 1.5\sigma$.  If an offset is confirmed, then it implies a complex emission feature, possibly indicative of a wind being launched from the vicinity of the LRD, similar to AGN-driven outflows measured in low-redshift ultraluminous infrared galaxies \citep{Rose_2018} and AGN \citep{Wylezalek_2020}.  However, in AGN-driven winds, the velocity offset of broad \oiii\ (what \citeauthor{Rose_2018} quantify as a ``trans-auroral blend'' of \oiii\ $\lambda$4364 and \hgamma) is the same offset as they broad \hbeta\ offset, which we do not observe, $\Delta \mathrm{v}(\hbeta) = 67 \pm 107$~km s$^{-1}$ (Table~\ref{table:optlines}).  Therefore, we disfavor that the broad \oiii\ $\lambda$4364 originates from an outflow of warm, dense gas, but this requires further study. 

\begin{figure}
    \centering
    \includegraphics[width=0.95\linewidth]{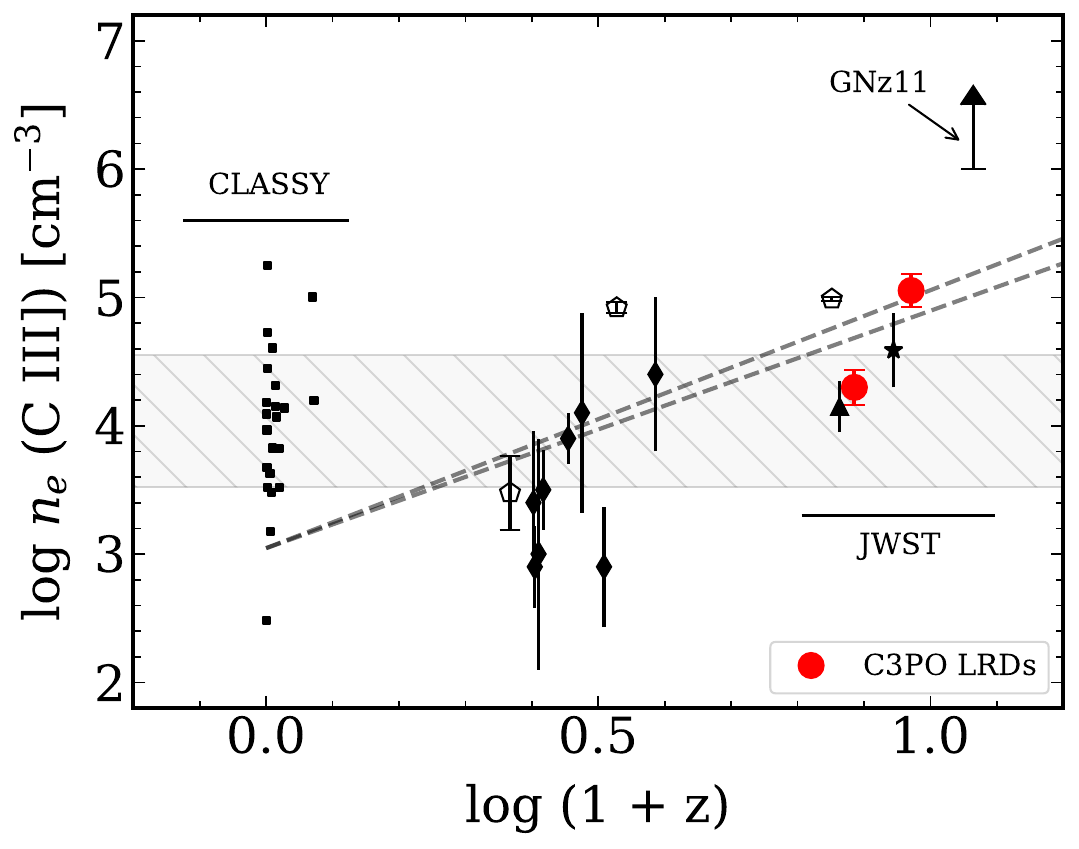}
    \caption{Gas density derived from the [\ion{C}{3}] $\lambda$1907 / \ciii\ $\lambda$1909 ratio for galaxies from the literature and from the C3PO LRDs here as a function of redshift.  The literature data include low-redshift galaxies from CLASSY \citep[circles]{Mingozzi_2022}, intermediate galaxies from the MUSE HUDF \citep[diamonds]{Maseda_2017}, and other high-redshift galaxies,  including individual well-measured values from \citet{Martinez_2025}, a median of $z\sim 6$ galaxies with \jwst\ data \citep[triangle,][]{Topping_2025},  and C3PO 13264 (star, C.~Papovich et al., in prep).  The C3PO LRDs are shown as large, red circles. The hatched region shows the inter-68\%-range for the CLASSY galaxies.  The dashed lines show the redshift evolution for $n_e(\ciii)$ from \cite{Martinez_2025}.    The The C3PO LRDs have high \ciii-derived electron densities, but not significantly different compared to the other high-redshift galaxies. The arrow shows the lower limit on the $n_e$ derived from the \ion{N}{4}] ratio for GNz11 \citep{Maiolino_2024_nature}, which is significantly higher than the constraints from \ciii\ for the LRDs here. }
    \label{fig:ciii_ne_z}
\end{figure}

\subsubsection{Structure of Gaseous Components}

The C3PO LRDs have high \ciii-derived electron densities, but they are not atypical for other galaxies with \ciii-based measurements. Figure~\ref{fig:ciii_ne_z} compares the gas densities using the [\ciii\ $\lambda$1907 / \ciii\ $\lambda$1909 line ratios for the C3PO LRDs to those derived from the same ratio for galaxies in the literature.  While the gas densities in the literature assume a range of gas temperature, $\approx 10,000$ to 20,000~K, the densities are relatively insensitive to this (see, Figure~\ref{fig:uvline_ratios}). The 16th--to--84th percentile range of low--redshift galaxies from CLASSY is $\log n_e/\mathrm{cm^{-3}} = 3.5-4.6$ \citep{Mingozzi_2022}. This covers $\log n_e = 4.3 \pm 0.3$ for C3PO 46403.  C3PO 45290 lies outside this range, with $\log n_e = 5.1 \pm 0.4$, but only by $\simeq 1.2\sigma$, and it falls in the range spanned by the full CLASSY sample.  The C3PO LRD \ciii--based densities are also consistent with other high-redshift measurements (Figure~\ref{fig:ciii_ne_z}).  In contrast, the measurement for GNz11 using \ion{N}{4}] shows much higher density, which is suggestive of regions associated with AGN \citep{Maiolino_2024_nature}. This implies the intermediate-/high-ionization zones of the narrow-line emitting clouds in the C3PO LRDs have gas densities similar to those of star-forming galaxies. 

\begin{figure}[t]
    \centering
    \includegraphics[width=\linewidth]{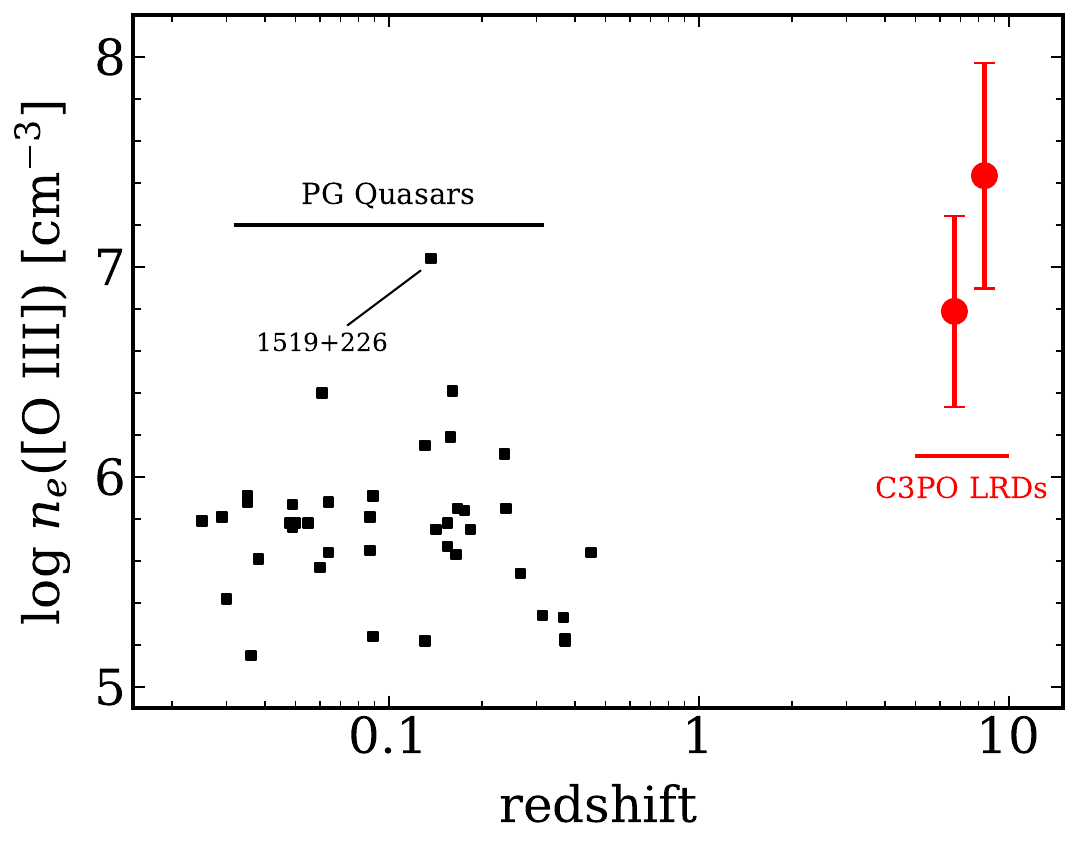}
    \caption{Gas densities derived from broad-line components of the \oiii\ $\lambda$4364 / \oiii\ $\lambda$5008 ratio for objects as a function of redshift.  The low-redshift points show results for PG Quasars \citep{Baskin_2005} derived from measurements of the broad components of the \oiii\ $\lambda$4364 and \oiii\ $\lambda$5008 emission lines.  The two C3PO LRDs are shown as large circles.  The gas densities in the broad-line regions of the LRDs are high compared to that in typical quasars. }
    \label{fig:blr_ne_z}
\end{figure}

In contrast, the gas densities in the broad \oiii\ emitting regions are high for the C3PO LRDs (Section~\ref{section:broad_gas}).   Figure~\ref{fig:blr_ne_z} compares the measurements for the gas densities derived from the broad \oiii\ $\lambda$4364 / \oiii\ $\lambda$5008 lines for the C3PO LRDs compared to those measured from the broad-line regions in PG Quasars \citep{Baskin_2005}.  The PG Quasars have gas densities that span $\log n_e(\oiii) / \mathrm{cm^{-3}} = 5.1$ to 7.0, with a interquartile range of 5.6--5.9.  In contrast, the C3PO LRDs have gas densities of their broad-lines of $\log n_e(\oiii) / \mathrm{cm^{-3}} = 6.3$ to 7.9, including their $1\sigma$ uncertainties (Table~\ref{table:quantities}).  C3PO 46403 appears to have a lower gas density than C3PO 45290, but this may result from \ion{He}{1} $\lambda$5017 emission contributing to the broad \oiii\ $\lambda$5008 component or attributing some of the \oiii\ $\lambda$4364 emission to [\ion{Fe}{2}] (see Section~\ref{section:broad_gas}).  %
%
Regardless, the gas densities measured within $\sim$10 pc of the central engine for these two LRDs are high compared to that of typical low-redshift quasars. 

The only PG quasar with a gas density as high as the C3PO LRDs is 1519+226 (labeled in Figure~\ref{fig:blr_ne_z}).  This object was noted as an outlier in \cite{Baskin_2005} and shows very weak \oiii\ $\lambda\lambda$4960,5008 emission \citep{Boroson_1992}.  The interpretation of PG 1519+226 is that it likely requires a dense structure physically located near or obscured by the dust torus in the classical AGN paradigm \citep{Nagao_2001}.  Notably, \citet{Nagao_2001} argue that AGN with high \oiii\ $\lambda$4364 / \oiii\ $\lambda$5008 also show stronger highly ionized emission from lines such as [\ion{Fe}{7}] $\lambda$6087.  While we do detect a line at the location of [\ion{Fe}{7}] $\lambda$5160 (see Figure~\ref{fig:46403_spec}), we do not detect other [\ion{Fe}{7}] lines, which should be present \citep[see,][]{Torralba_2026}.  For example, using \texttt{PyNeb}, the ratio [\ion{Fe}{7}] $\lambda$6087/[\ion{Fe}{7}] $\lambda$5160 is  $\geq 3$. Our $3\sigma$ upper limit on this ratio is [\ion{Fe}{7}] $\lambda$6087/[\ion{Fe}{7}] $\lambda$5160 $<$ 0.7, calling this identification into question (and may result from a blend of [\ion{Fe}{2}] lines, see also \citealt{Lambrides_2026,Torralba_2026}).  Regardless, both C3PO LRDs here show strong narrow \oiii\ $\lambda\lambda$4960,5008, indicating different conditions exist in them compared to PG 1519+226, or to the other PG quasars. 



\subsection{Constraints on the Nature of LRDs} 


Our results suggest that most, but not all, of the ionizing UV emission from an LRD accretion disk is obscured by the LRD gaseous envelope.  This is required to explain the UV emission-line ratios of \civ/\ciii, \oiiiuv/\heii, and \ciii/\heii\ (Section~\ref{section:ionzing_source} and Figure~\ref{fig:uvline_diagnostics}), which favor ionization conditions from stars with some  contribution from an accretion disk.  This is consistent with the densities we measure from the narrow \ciii--emission lines, which indicate gas densities similar to other star-forming galaxies at low and high redshifts (Section~\ref{section:discussion_structure} and Figure~\ref{fig:ciii_ne_z}).  The narrow emission-line ratios indicate that star-formation contributes, or even dominates their ionization.  

Two more pieces of evidence support the assertion that the emission lines require some ionizing radiation from the accretion disk of the LRD.  The first piece comes from the \oiii\ $\lambda$4364 / \hgamma\ ratios, where ionization from an accretion disk is favored based on the ratio of \oiii\ $\lambda$4364/\hgamma\ for the broad-line components in C3PO 45290, and for the ratio of the total or narrow--line components in C3PO 46403 (Section~\ref{section:ionzing_source} and Figure~\ref{fig:o3hg}).  The second piece comes from the EW of the UV \ciii\  and \oiiiuv\ lines in both C3PO LRDs, which are larger than possible from star-formation alone (Section~\ref{section:ionzing_source} and Figure~\ref{fig:uvline_diagnostics}).  This implies that some ionizing flux from the LRD accretion disk escapes into both the broad-line region and the narrow-line regions.  

The C3PO LRDs therefore favor a scenario where the gaseous envelopes of the LRD accretion disk have a non-unity covering factor, where the broad \oiii\ $\lambda$4364 traces high-density clouds within $\sim$10~pc of the LRD center.  The non-unity covering factor is required such that some ionizing radiation from the LRD accretion is necessary to explain the LRD emission--line properties.  For these two LRDs, this disfavors models of the gaseous envelopes that require unity covering factors \citep[e.g.,][]{Naidu_2025}.   Our finding that the covering factors in our LRDs are less than unity seems at odds with the argument that the shape of the broad-lines stems from electron scattering, as these require a near-spherical gas distribution, else the scattered light preferentially escapes along the lowest column-density sightlines \citep{Rusakov_2026}.  Intriguingly, our results may support the scenarios proposed by \citet{Davis_2026} based on measurements of Balmer-absorption lines in LRDs using high-resolution \jwst\ NIRSpec observations.  In this scenario, the Balmer-absorption originates from smaller, distinct clouds, rather than a uniform shell, and that these are distributed radially around the LRD accretion disk.  In this model the clouds are self-shielding, such that broad emission lines form on the side of the cloud facing the accretion disk, while the shielded side produces absorption.  This is similar to the stratified gas-envelope model posited by \citet{Madau_2026c}, and consistent with the observations of \citet{Tang_2026} for a clumpy gas envelope based on the properties of Lyman-$\alpha$ emission in LRDs.  Our observations support this model as it includes non-unity covering fractions, such that ionizing radiation from the LRD is able to contribute to the broad \oiii\ $\lambda$4364 and partially to the narrow UV \ciii\ and \oiiiuv\ lines.  In this model, the broad \oiii\ $\lambda$4364 lines would trace kinematics at distances given by Equation~\ref{equation:radius}, or 1--10~pc from the LRD engine (Section~\ref{section:discussion_structure}).  Even at these distances the clouds responsible for the broad \oiii\ $\lambda$4364 emission have densities $>3\times 10^{6}$~cm$^{-3}$ for both C3PO LRDs. 

Our observations are consistent with scenario where the LRD engine is powered by accretion onto a SMBH surrounded by stratified layers of gas clouds.  This can be consistent with a quasi-star model theorized by \citet[see also, \citealt{Begelman_2026}]{Begelman_2008}, but with a non-unity covering factor.  In this model, an accreting SMBH is embedded within a  gaseous envelope in hydrostatic equilibrium, where the accretion rate is able to exceed the Eddington limit of the SMBH itself.  The gas thermalizes the emission from the accretion, with effective temperatures of $\sim$4000--5000~K, satisfying observations of the SED shape of LRDs.   The basic quasi-star model assumes spherical symmetry and unity covering factors, but this is not required:  \citet{Begelman_2008} suggest that real situations can include gas--envelope flattening from rotation and accretion-disk formation around the SMBH, where complex geometries are possible.  This is similar to the ``blackbody'' model for LRDs favored by \citet{Asada_2026}.    For the quasi-star model to match our observations, complex geometries seem required to allow spatial variations to occur, where the radiation pressure can exceed gravity and clear sightlines through which ionizing radiation from the accretion disk can escape to contribute to ionization of the broad \oiii\ $\lambda$4364 region and narrow UV emission lines.   In this case our observations here provide an important measurement that this situation occurs within this model. 

One challenge to the quasi-star interpretation is that the escaping radiation from the LRD accretion disk must be sufficient to produce photons that contribute to the production of O$^{2+}$ in the broad-line region (ionization potential of 35.1 eV) and contribute to the production of O$^{2+}$ and C$^{2+}$ (ionization potential of 24.4~eV) in the narrow-line regions, without contributing to the ionization of He$^{2+}$ (creation energy of 54.4 eV) or higher energy lines.  If the shape of the the ionizing radiation of the accretion disk matches expectations for lower-mass SMBHs, we may expect harder spectra, producing lower \ciii\ $\lambda$1907/\heii\ $\lambda$1640, \oiiiuv\ $\lambda$1666/\heii\ $\lambda$1640 (Figure~\ref{fig:uvline_diagnostics}), or even [\ion{Ne}{5}]/[\ion{Ne}{3}] \citep{Feltre_2016,Hirschmann_2019,Cleri_2023b}, which we do not observe.  Similarly, studies of \heii\ $\lambda$4687/\hbeta\ argue for a lack of hard ionizing photons from the LRD \citep{Sok_2026}. 


To explain the reduced \heii\ $\lambda$1640 emission, the intrinsic ionizing spectrum of the LRD is likely inconsistent with that of standard AGN.  This is consistent with the findings of  \citep{Wang_2026}. A physical mechanism to explain this may be similar to that studied by \citet{Laor_2014}, who argue that line-blanketing is able to drive winds from AGN accretion disks that are able to produce sufficient mass loss to keep the effective temperature of the disk $\ll 10^5$~K, similar to the limiting process in O-stars. These models predict a cold AGN accretion disk with a turnover in the SED at $\lambda >500$~\AA, reducing the number of hard photons at $E \gtrsim 25$~eV, qualitatively consistent with the observations here.   Therefore, the necessity to suppress hard ionizing photons in the accretion disks of LRDs may favor models where winds or other physical processes produce intrinsically colder SEDs in LRDs. 

Lastly, we stress that our interpretation of the gaseous envelope of C3PO LRDs here may not extend to the full population.  It is likely, even probable, that LRDs represent a transitory phase in the growth of SMBHs at these redshifts.  Indeed, one hallmark of most LRD models is that their gaseous envelopes are not in hydrostatic equilibrium, but supported by radiation pressure \citep{Begelman_2026}, which are inherently unstable.  Therefore, while our observations here are consistent with a stratified gaseous envelope and/or quasi-star--like model, other LRDs may be found in different stages, with different physical conditions \citep[see, e.g.,][]{Naidu_2025}.  

%
%
%


\section{Conclusions}\label{section:summary}

We presented deep NIRSpec spectroscopy in G140M and G395M of two LRDs from the C3PO survey, C3PO 46403 at $z=6.68$, and C3PO 45290 at $z=8.35$.  The spectroscopy cover important diagnostic emission features in the rest-frame UV and optical.   We used these data to constrain the ionization, densities, and kinematics in these two LRDs.    Our conclusions are as follows.  

Both LRDs show broad Balmer and \oiii\ $\lambda$4364 emission in the G395M data.  This is the first report of broad \oiii--emission lines in LRDs.  The flux ratios of the broad \oiii\ $\lambda$4364/\oiii\ $\lambda$5008 indicate high gas densities, $\log n/\mathrm{cm^{-3}} = 6.3$ to 7.9. These are higher by a factor of 3--10 than those in broad-line regions of typical low-redshift quasars.  The broad \oiii\ $\lambda$4364 lines have FWHM~$\simeq$1000~km s$^{-1}$, about one--third of the width of the broad \hbeta\ lines.  If these trace virial motions, then this is evidence for metal-enhanced, dense gas clouds, $\sim$1-10~pc from the LRD engine.  
%
%

Both LRDs show narrow UV metal lines in the G140M data, including [\ion{C}{3}] $\lambda$1907 + \ciii\ $\lambda$1909, and \oiiiuv\ $\lambda\lambda$1661,1666.  The \ciii/\heii\ and \oiiiuv/\heii\ ratios are generally consistent with ionization from stellar populations with lower metallicity and lower C/O ratios.  The \ciii\ line ratios yield gas densities, $\log n_e/\mathrm{cm^{-3}} = 4.2-5.2$, which are high, but not atypical of \ciii--based densities in star-forming galaxies at low and high redshifts.  We also detect other density--sensitive emission lines in the UV, \ion{Si}{3}] and \ion{N}{4}], at weaker significance, but which provide density constraints consistent with the \ciii--based measurements. 

The UV emission line ratios, \civ/\ciii, \ciii/\heii, and \oiii/\heii, are consistent with model predictions for ``composite'' objects that can contain either star-formation or AGN.   However, the UV emission-line ratios are consistent with other extreme star-forming objects at low and high redshift, indicating that AGN are not required, and that star-formation is a likely dominant source of ionization in these lines. 

However, there are two pieces of evidence that some contribution to the ionization from an AGN is necessary in both LRDs studied here.  First, the UV line equivalent widths, EW(\oiiiuv), EW(\ciii), are at, or exceed, limits expected for stellar populations, and require an additional source of ionization.    Second, the \oiii\ $\lambda$4364/\hgamma\ emission line ratios favor ionization from an AGN combined with stars.  Therefore, some ionizing radiation from the LRD accretion disk seems required to explain the emission lines.   However, this places an intriguing limit on the intrinsic shape of the ionizing spectrum of the LRD accretion disk as it must provide photons to contribution to the production O$^{2+}$ and C$^{2+}$ (ionization potentials of 35.1 and 24.4~eV, respectively) without contributing to the production of He$^{2+}$ (ionization potential of 54.4~eV) or higher energy lines.  This may be possible with line-driven winds from the LRD accretion disk, similar to the process in O-stars \cite{Laor_2014}, but this requires further study.

These results favor a scenario where the LRD gas envelopes are highly stratified, having high-density clouds with a non-unity covering factors with a complex geometry, such that ionizing radiation from the LRD, combined with that from star-forming regions, produce the nebular emission features. 

Both LRDs show tentative evidence of low oxygen abundances based on analysis of their rest-optical emission-line ratios.  The LRDs also show evidence for sub-solar C/O ratios based on their rest--UV emission line ratios.  Interestingly, both LRDs also show evidence of N/O enhancement based on detections of either \ion{N}{3}] $\lambda\lambda$1746,1748 in C3PO 46403, or \ion{N}{4}] $\lambda$1486 in C3PO 45290.   These elemental abundance ratios could be evidence for rapid and recent star-formation, in particular to produce the elevated N/O ratios.   If nitrogen lines and high N/O ratios are common among LRDs, this means that short, intense starbursts coincide or immediately precede the LRD phase. This could lead to a nature explanation that dense gas in galaxies drives both intense star formation and the formation of an LRD.  

\begin{acknowledgements}

This work benefited from support from the George P. and Cynthia Woods Mitchell Institute for Fundamental Physics and Astronomy at Texas A\&M University. CP thanks Marsha and Ralph
Schilling for generous support of this research.  
%
%
This work is based on observations made with the NASA/ESA/CSA James Webb Space Telescope. The data were obtained from the Mikulski Archive for Space Telescopes at the Space Telescope Science Institute, which is operated by the Association of Universities for Research in Astronomy, Inc., under NASA contract NAS 5-03127 for JWST. These observations are associated with program \#5943.  Support for US investigators in program \#5943 was provided by NASA through a grant from the Space Telescope Science Institute, which is operated by the Association of Universities for Research in Astronomy, Inc., under NASA contract NAS 5-03127.

\end{acknowledgements}

The data described here may be obtained from the MAST archive at \dataset[doi:10.17909/wey4-5s43]{https://dx.doi.org/10.17909/wey4-5s43}.

\facility{JWST (NIRSpec)}

\software{ \texttt{Astropy} \citep{Astropy_2013}, \texttt{Specutils} \citep{astropy_specutils},
JWST Calibration Pipeline \citep{Bushouse_2025}, \texttt{Cloudy} \citep{cloudy25}, \texttt{PyNeb} \citep{Luridian_2015}
}

\bibliographystyle{aasjournalv7}
\bibliography{c3po_BLR}{}

\end{document}